\documentclass[nofootinbib,superscriptaddress,amssymb,prd,aps,amsmath,twocolumn,preprintnumbers]{revtex4}

\usepackage{color}
\usepackage{graphicx}
\usepackage{hyperref}
\hypersetup{colorlinks,bookmarksopen,bookmarksnumbered,citecolor=blue,
linkcolor=black,pdfstartview=FitH,urlcolor=blue}
\usepackage{multirow}
\usepackage[utf8]{inputenc}
\usepackage{url}
\usepackage{ulem}
\begin{document}

\preprint{TUM-HEP 946/14}
\preprint{IPPP/14/48}
\preprint{DCPT/14/96}

\title{Sharp Gamma-ray Spectral Features from Scalar Dark Matter Annihilations}

\author{Alejandro Ibarra}
\email{alejandro.ibarra@ph.tum.de}
\affiliation{Physik-Department T30d, Technische Universit\"at M\"unchen,
 James-Franck-Stra\ss{}e, 85748 Garching, Germany}

\author{Takashi Toma}
\email{takashi.toma@durham.ac.uk}
\affiliation{Institute for Particle Physics Phenomenology University of
Durham,  Durham DH1 3LE, United Kingdom}

\author{Maximilian Totzauer}
\email{maximilian.totzauer@mytum.de}
\affiliation{Physik-Department T30d, Technische Universit\"at M\"unchen,
 James-Franck-Stra\ss{}e, 85748 Garching, Germany}

\author{Sebastian Wild}
\email{sebastian.wild@ph.tum.de}
\affiliation{Physik-Department T30d, Technische Universit\"at M\"unchen,
 James-Franck-Stra\ss{}e, 85748 Garching, Germany}

\begin{abstract}
The search for sharp features in the gamma-ray spectrum is a promising approach to identify a signal from dark matter annihilations over the astrophysical backgrounds. In this paper we investigate the generation of gamma-ray lines and internal bremsstrahlung signals in a toy model where the dark matter particle is a real scalar that couples to a lepton and an exotic fermion via a Yukawa coupling. We show that the Fermi-LAT and H.E.S.S. searches for line-like spectral features severely constrain regions of the parameter space where the scalar dark matter is thermally produced. Finally, we also discuss the complementarity of the searches for sharp spectral features with other indirect dark matter searches, as well as with direct and collider searches.
\end{abstract}

\date{\today}

\pacs{}
\keywords{}

\maketitle

\section{Introduction}

Multiple astrophysical and cosmological observations have demonstrated that a significant fraction of the matter content of the Universe is in the form of new particles not included in the Standard Model, but belonging to the so-called dark sector (see \cite{Bertone:2004pz,Bergstrom:2000pn} for reviews). The dark matter (DM) particles were presumably produced during the very early stages of the Universe and must have a relic abundance today $\Omega_{\rm DM} h^2\approx0.12$~\cite{Ade:2013zuv}. Among the various production mechanisms proposed, the freeze-out mechanism stands among the most appealing and predictive ones. In this framework, the dark matter particles were in thermal equilibrium with the Standard Model particles at very early times, but went out of equilibrium when the temperature reached a value $\sim m_{\rm DM}/25$. Below this temperature, the expansion rate became larger than the annihilation rate and therefore the number density of dark matter particles per comoving volume remained practically constant until today, the value being inversely proportional to their annihilation cross section into Standard Model particles. 

The annihilations that lead to the freeze-out of dark matter particles in the early Universe presumably continue today, at a much smaller rate, in regions with high dark matter density, such as in galactic centers. There exists then the possibility of testing the freeze-out mechanism if the flux of energetic particles produced in the annihilations is detected at the Earth. Unfortunately, the expected flux from annihilations is typically much smaller than the background fluxes from astrophysical processes, which makes a potential signal difficult to disentangle from the still poorly understood backgrounds. 

A promising strategy to identify a dark matter signal is the search for sharp gamma-ray spectral features, such as gamma-ray lines~\cite{Srednicki:1985sf,Rudaz:1986db,Bergstrom:1988fp}, internal electromagnetic bremsstrahlung~\cite{Bergstrom:1989jr,Flores:1989ru,Bringmann:2007nk} or gamma-ray boxes~\cite{Ibarra:2012dw}. Most dark matter models predict rather faint sharp spectral features, however, the predicted signatures are qualitatively very different to the ones expected from known astrophysical processes, thus allowing a very efficient background subtraction. As a result, searches for sharp gamma-ray spectral features provide limits on the model parameters which are competitive, and sometimes better, than those from other approaches to indirect dark matter detection. 

Recent works have thoroughly investigated the generation of sharp gamma-ray spectral features in simplified models, as well as the complementarity of the searches for spectral features with other search strategies, in scenarios where the dark matter particle is a Majorana fermion that couples to a Standard Model fermion via a Yukawa coupling \cite{Bringmann:2012vr,Garny:2013ama,Chang:2013oia,An:2013xka,Bai:2013iqa, DiFranzo:2013vra,Kopp:2014tsa,Garny:2014waa}, or in the inert doublet dark matter model~\cite{Gustafsson:2007pc,LopezHonorez:2006gr,Garcia-Cely:2013zga}.

In this paper, we investigate the generation of sharp gamma-ray spectral features in the toy model of real scalar dark matter considered in ref.~\cite{Toma:2013bka, Giacchino:2013bta}, where the Standard Model is extended with a real singlet scalar $\chi$, candidate for dark matter, and an exotic vector-like fermion $\psi$, which mediates the interactions with the Standard Model fermions. In this model, the stability of the dark matter particle is ensured by imposing a discrete $\mathbb{Z}_2$ symmetry, under which $\chi$ and $\psi$ are odd while the Standard Model particles are even.  We assume for simplicity that the new sector only couples to a right-handed lepton of one generation, $f_R=e_R$, $\mu_R$ or $\tau_R$, in order to suppress potential contributions to lepton flavor violating processes such as $\mu\rightarrow e \gamma$. Under these simplifying assumptions the interaction Lagrangian of the dark matter particle with the Standard Model particles reads:
\begin{equation}
-{\cal L}_{\rm int}=\frac{\lambda}{2}\chi^2(H^\dagger H)+(y \chi \bar \psi f_R+{\rm h.c.})
\end{equation}
where $H$ is the Standard Model Higgs doublet. 

This model has the peculiarity that the  cross section for the tree-level two-to-two annihilation process $\chi \chi\rightarrow f\bar f$, which sets the relic abundance over large regions of the parameter space,  is d-wave suppressed in the limit $m_f\rightarrow 0$. On the other hand, the processes generating gamma-ray lines at the one loop level $\chi\chi\rightarrow \gamma \gamma,\gamma Z$ or internal bremsstrahlung $\chi\chi\rightarrow f\bar f \gamma$ proceed in the s-wave. Therefore, for values of the parameters leading to the correct relic abundance, the expected indirect detection signals are relatively large compared to other models and, under some conditions, at the reach of present instruments~\cite{Toma:2013bka, Giacchino:2013bta}. 

The paper is organized as follows. In Section \ref{sec:gamma-ray} we present the result for the cross sections and we discuss the relative strength of both signals. In Section \ref{sec:constraints} we present constraints on the model from perturbativity, thermal production, direct detection, indirect detection with charged cosmic rays  and collider experiments. In Section \ref{sec:numerics} we present a numerical analysis showing the complementarity of all these constraints, under the assumption that the dark matter particle was thermally produced. Finally, in Section \ref{sec:conclusions} we present our conclusions.

\section{Sharp gamma-ray spectral features from scalar dark matter annihilations}
\label{sec:gamma-ray}

The gamma-ray flux generated by the annihilation of scalar dark matter particles receives several contributions. In this paper we will concentrate on the generation of sharp gamma-ray spectral features, which, if observed, would constitute a strong hint for dark matter annihilations. We will neglect, however, the gamma-ray emission generated by the inverse Compton scattering of the electrons/positrons produced in the annihilation on the interstellar radiation field, and will only briefly discuss the gamma-rays produced by the decay and hadronization of Higgs or gauge bosons since they do not generate sharp features in the gamma-ray spectrum.

\begin{figure*}[t]
\begin{center}
\includegraphics[scale=0.8]{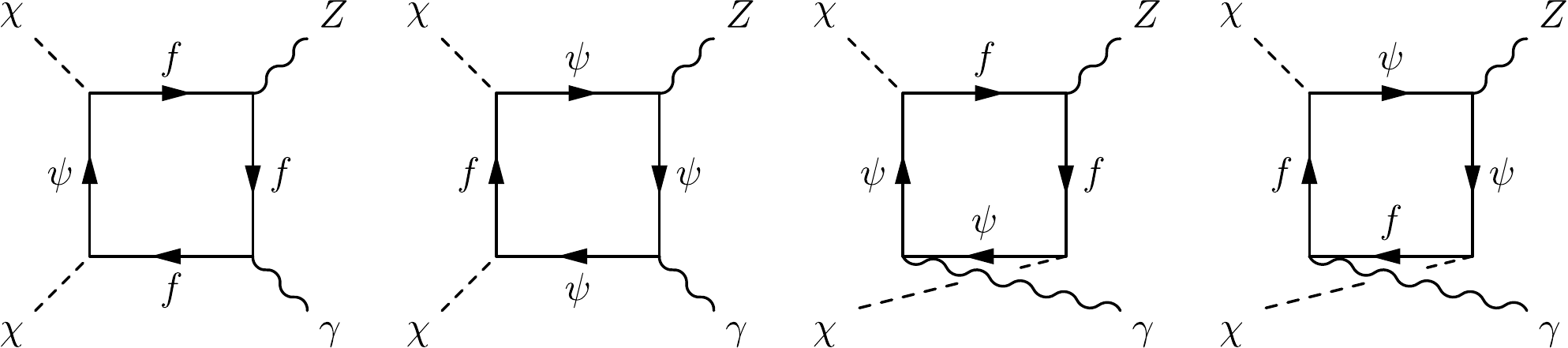}
\caption{Feynman diagrams inducing the annihilation $\chi\chi\to \gamma Z$ at the one loop level. The diagrams generating the annihilation into $\gamma\gamma$ can be obtained by replacing the $Z$-boson in the final state by a photon. Equivalent topologies with crossed initial or final state legs are not shown here.}
\label{fig:box}
\end{center}
\end{figure*}

The scalar $\chi$ does not have tree-level electromagnetic interactions. Nevertheless, annihilations into one or two photons are possible via higher order effects mediated by the Standard Model fermion $f$ and the exotic fermion $\psi$, which do carry electroweak charges. More specifically, the interaction Lagrangian with photons of the charged fermions of the model reads:
\begin{equation}
\mathcal{L}_{\mathrm{QED}}=eA_{\mu}\overline{f}\gamma^{\mu}f
+eA_{\mu}\overline{\psi}\gamma^{\mu}\psi, 
\end{equation}
while the interaction Lagrangian with the $Z$ boson reads:
\begin{equation}
\mathcal{L}_{Z}=
- e \tan\theta_W  Z_{\mu}\overline{f_R}\gamma^{\mu} f_R
-e \tan\theta_W  Z_{\mu}\overline{\psi}\gamma^{\mu}\psi.
\end{equation}\pagebreak

The annihilations $\chi\chi\rightarrow \gamma\gamma$ and $\gamma Z$ are generated at the one loop level, through the diagrams shown in Fig.~\ref{fig:box}. In the Milky Way center, dark matter particles are expected to be very non-relativistic, $v\approx 10^{-3}$, thus generating a monoenergetic photon in the annihilation process. In the limit of zero relative velocity, the transition amplitude for the annihilation $\chi\chi\rightarrow \gamma\gamma$ can be cast as
\begin{equation}
i\mathcal{M}_{\gamma\gamma}=-\frac{i\alpha_{\mathrm{em}}y^2}{\pi}
\epsilon_{\mu}^*(k_1)\epsilon_{\nu}^*(k_2) g^{\mu\nu} \mathcal{A}_{\gamma\gamma},
\label{eq:MGammaGamma}
\end{equation}
as required by gauge invariance. Here,  $\epsilon_\mu(k)$ is the polarization vector of the photon, 
$\alpha_{\mathrm{em}}$ is the electromagnetic fine structure
constant, defined as $\alpha_{\mathrm{em}}\equiv e^2/(4\pi)$ and
$\mathcal{A}_{\gamma\gamma}$ is a form factor. The explicit expression
for the form factor $\mathcal{A}_{\gamma\gamma}$  at the one loop level
is rather complicated and is reported in the Appendix. The form factor
greatly simplifies in the limit $m_f\to0$ and reads\footnote{In~\cite{Boehm:2006gu}, the $\gamma \gamma$ amplitude was calculated for a scenario of MeV dark matter, but for the case of couplings to both left- and righthanded fermions, to leading order in $1/m_\psi$.}: 
\begin{equation}
\mathcal{A}_{\gamma\gamma}=
2+\mathrm{Li}_2\left(\frac{1}{\mu}\right)-\mathrm{Li}_2\left(-\frac{1}{\mu}\right)
-2\mu\,\mathrm{arcsin}^2\left(\frac{1}{\sqrt{\mu}}\right),
\label{eq:loop-f}
\end{equation}
with $\mu\equiv m_\psi^2/m_\chi^2$. \footnote{Our result differs from the one reported in ~\cite{Bertone:2009cb,Tulin:2012uq}. In particular, we obtain a finite amplitude when $\mu\rightarrow 1$.} Finally, the annihilation cross section for $\chi\chi\rightarrow \gamma\gamma$ is given by 
\begin{equation}
\sigma{v}_{\gamma\gamma}=\frac{y^4\alpha_{\mathrm{em}}^2}
{32\pi^3 m_\chi^2}\left|\mathcal{A}_{\gamma\gamma}\right|^2.
\label{eq:sigmatotal2gamma}
\end{equation}

On the other hand, the transition amplitude for $\chi\chi\to\gamma Z$
can be cast, in the zero velocity limit, as 
\begin{equation}
i\mathcal{M}_{\gamma Z}=\frac{i y^2\alpha_{\mathrm{em}}\tan\theta_W}{\pi}
\epsilon_{\mu}^*(k_1)\epsilon_{\nu}^*(k_2)g^{\mu\nu} \mathcal{A}_{\gamma
Z} ,
\label{eq:MGammaZ}
\end{equation}
where $\epsilon_\mu(k_1)$, $\epsilon_\nu(k_2)$ are the polarization vectors of the $Z$ boson and the photon, respectively, and $\mathcal{A}_{\gamma Z}$ is the corresponding form factor, which is reported in the Appendix. The cross section is in this case given by 
\begin{equation}
\sigma{v}_{\gamma Z}=\frac{y^4 \alpha_{\mathrm{em}}^2\tan^2\theta_W}{16\pi^3
 m_\chi^2}
 \left(1-\frac{m_Z^2}{4m_\chi^2}\right)\left|\mathcal{A}_{\gamma Z}\right|^2.
\label{eq:sigmatotalZgamma}
\end{equation}

\begin{figure*}[t]
\begin{center}
\includegraphics[scale=0.8]{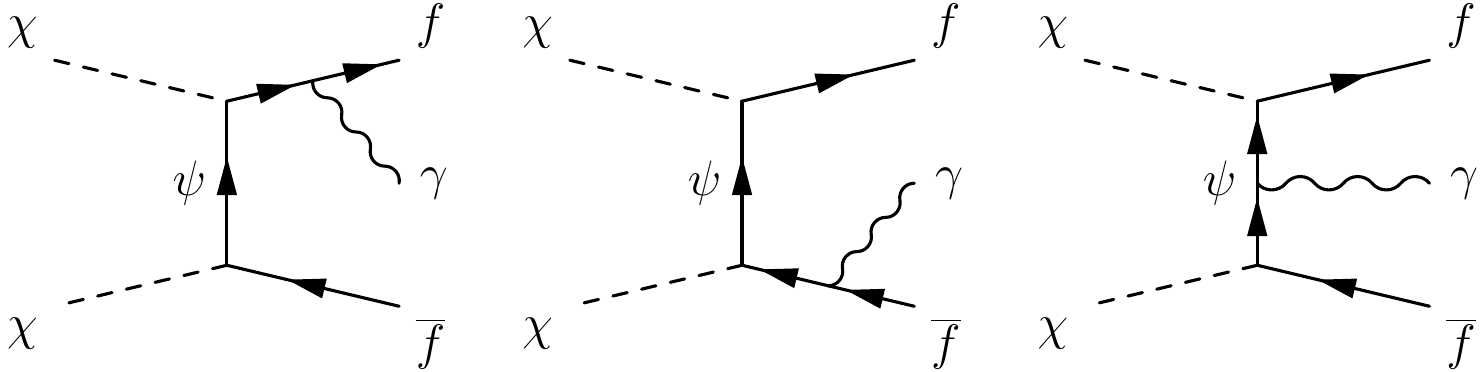}
\caption{Feynman diagrams inducing the internal bremsstrahlung process $\chi\chi\to f\overline{f}\gamma$.}
\label{fig:vib}
\end{center}
\end{figure*}

On the other hand, the two-to-three annihilation into a fermion-antifermion pair with the associated emission of a photon results from Feynman diagrams shown in Fig.~\ref{fig:vib}, where the photon can be attached to either of the charged fermions. The differential cross section for this process reads:\pagebreak
\begin{widetext}
\begin{equation}
\frac{d\sigma{v}_{f\overline{f}\gamma}}{dx}=
\frac{y^4\alpha_{\mathrm{em}}}{4\pi^2 m_\chi^2}\left(1-x\right)
\left[
\frac{2x}{\left(\mu+1\right)\left(\mu+1-2x\right)}
-\frac{x}{\left(\mu+1-x\right)^2}
-\frac{\left(\mu+1\right)\left(\mu+1-2x\right)}{2\left(\mu+1-x\right)^3}
\log\left(\frac{\mu+1}{\mu+1-2x}\right)
\right], 
\end{equation}
with $x=E_\gamma/m_\chi$, while the total cross section is given by 
\begin{equation}
\sigma{v}_{f\overline{f}\gamma}=
\frac{y^4\alpha_{\mathrm{em}}}{8\pi^2 m_\chi^2}
\left[
\left(\mu+1\right)\left\{\frac{\pi^2}{6}-\log^2\left(\frac{\mu+1}{2\mu}\right)
-2\mathrm{Li}_2\left(\frac{\mu+1}{2\mu}\right)\right\}
+\frac{4\mu+3}{\mu+1}+\frac{\left(4\mu+1\right)\left(\mu-1\right)}{2\mu}
\log\left(\frac{\mu-1}{\mu+1}\right) 
\right].
\label{eq:sigmatotal2to3}
\end{equation}
\end{widetext}
As it is well known,  the gamma-ray spectrum for the two-to-three process displays a sharp peak close to the kinematical endpoint of the spectrum, which becomes more and more prominent as $\mu\rightarrow 1$~\cite{Bringmann:2007nk}.

\begin{figure}[t]
\begin{center}
\includegraphics[scale=0.65]{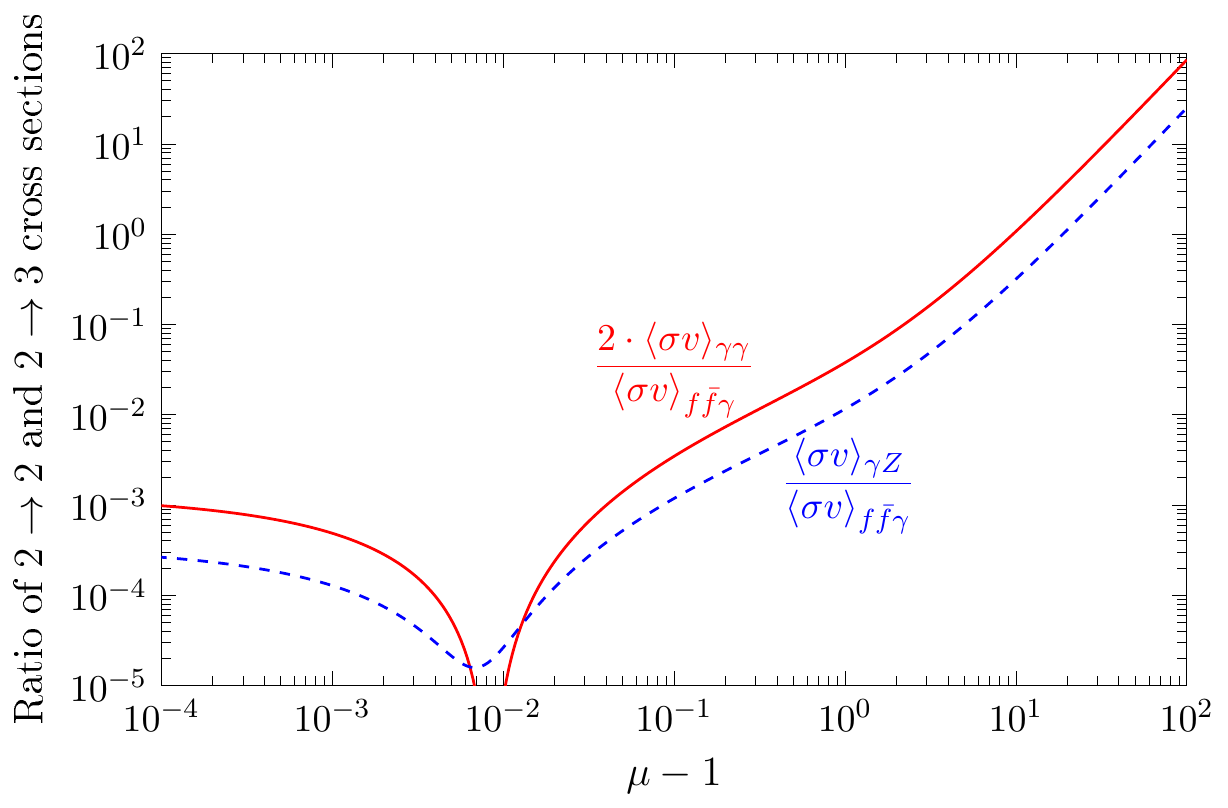}
\caption{Ratio between the cross sections of the one-loop induced two-to-two annihilation into $\gamma\gamma$ (or $\gamma Z$) and the two-to-three annihilation into $f\bar f\gamma$, as a function of the parameter $\mu\equiv m_\psi^2/m_\chi^2$. For the plot it was assumed $m_\chi = 500$ GeV.}
\label{fig:ratio}
\end{center}
\end{figure}

The relative importance of the one loop processes $\chi \chi\rightarrow \gamma V$, with $V=\gamma, Z$, and the three body process $\chi \chi\rightarrow f\bar f \gamma$ is determined by $\mu\equiv m_\psi^2/m_\chi^2$ and by $m_V^2/m_\chi^2$.  This dependence is explicitly shown in Fig.~\ref{fig:ratio}, where we have taken $m_\chi=500$ GeV for definiteness. For $\mu=1$ the cross section for the two-to-three process is $1.6 \times 10^3$ ($2.9 \times 10^3$) times larger than the cross section for $\gamma\gamma$ ($\gamma Z$).  As $\mu$ increases, $|\mathcal{A}_{\gamma \gamma}|$ decreases and eventually changes sign at $\mu-1 \simeq 8.0 \times 10^{-3}$. Accordingly, at this point the $\gamma \gamma$ cross section vanishes. For larger mass splittings, the relative importance of the one loop processes increases and they become the dominant process when $\mu\gtrsim 10$ ($\gtrsim 19$). A similar behavior arises in scenarios with Majorana dark matter particles coupling to leptons via a Yukawa coupling with a scalar field, as discussed in ref.~\cite{Garny:2013ama}.

\begin{figure*}[t]
\begin{center}
\includegraphics[scale=0.9]{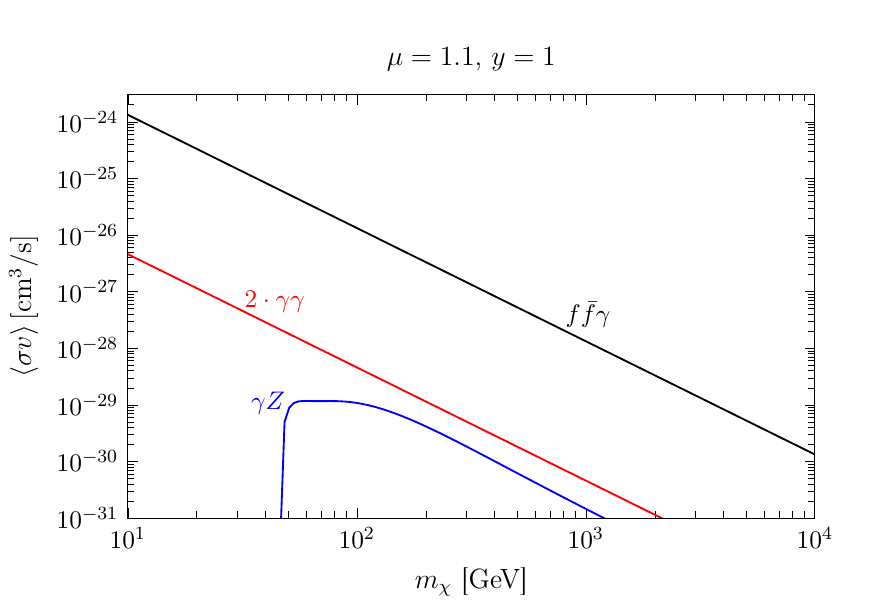}
\hspace{0.2cm}
\includegraphics[scale=0.9]{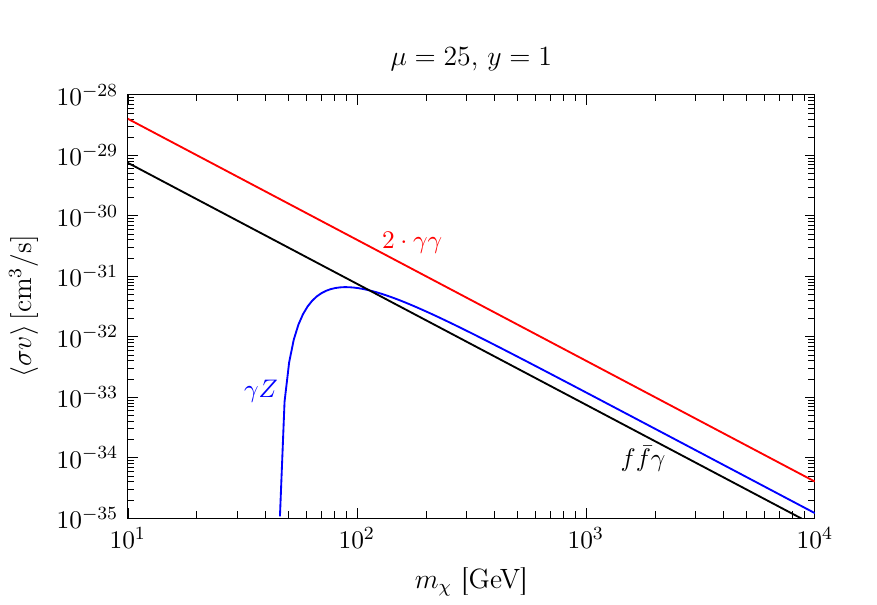}
\\
\includegraphics[scale=0.9]{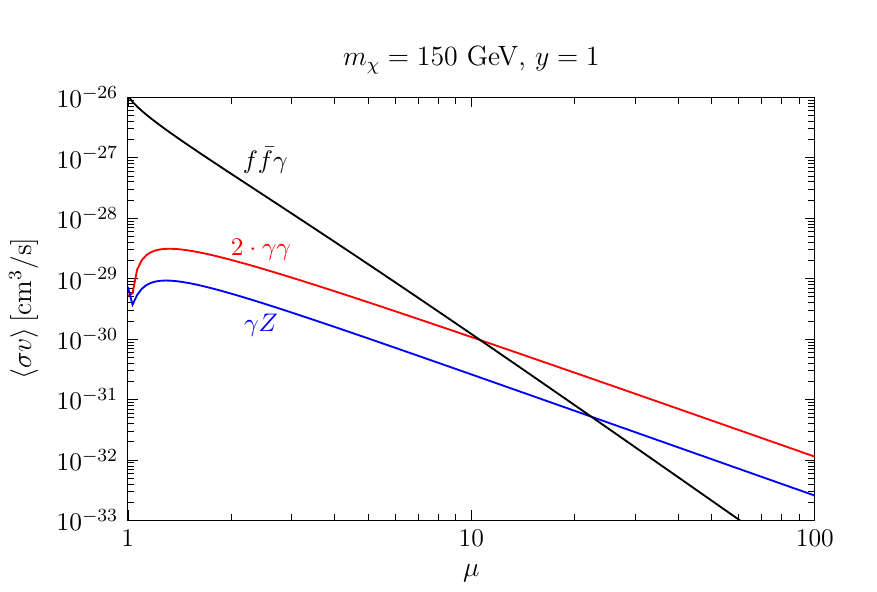}
\hspace{0.2cm}
\includegraphics[scale=0.9]{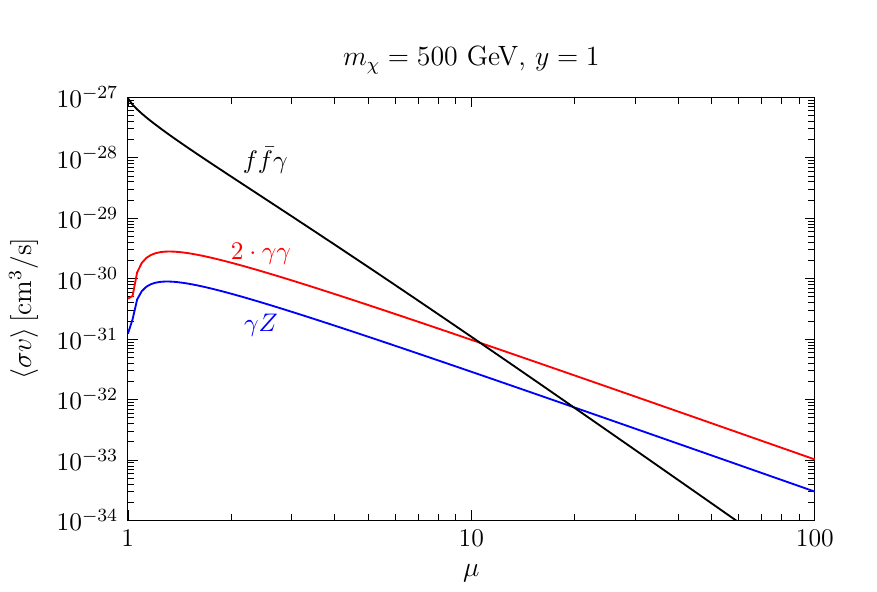}
\end{center}
\caption{Annihilation cross section for the processes $f\bar f \gamma$ (black line), $\gamma\gamma$ (red line) and $\gamma Z$ (blue line) for some exemplary choices of parameters. The top plots show the cross sections as a function of the dark matter mass for $\mu=1.1$ (left plot) and $\mu=25$ (right plot), while the bottom plots show the cross sections as a function of the parameter $\mu\equiv m_\psi^2/m_\chi^2$ for $m_\chi=150$ GeV (left plot) and $m_\chi=500$ GeV (right plot). All plots assume $y=1$. }
\label{fig:vib_vs_loop}
\end{figure*}

The gamma-ray flux generated by the two-to-three and the one loop annihilation processes depends on the cross sections in those channels. The dependence of these cross sections on the different parameters of the model is shown in Fig.~\ref{fig:vib_vs_loop}, where we have fixed in all the plots the Yukawa coupling $y$ to 1. Approximate  analytic expressions for the cross sections when $\mu\gg 1$ are
\begin{align*}
\langle \sigma v \rangle_{f \bar{f} \gamma} &\simeq 3.0 \times 10^{-26} \, \text{cm}^3\text{s}^{-1} \frac{y^4}{\mu^4}  \left( \frac{100 \,\text{GeV}}{m_\chi} \right)^2  \nonumber\\
\langle \sigma v \rangle_{\gamma \gamma} &\simeq 1.3 \times 10^{-28} \, \text{cm}^3\text{s}^{-1} \frac{y^4}{\mu^2}  \left( \frac{100 \,\text{GeV}}{m_\chi} \right)^2 \,, \nonumber\\
\langle \sigma v \rangle_{\gamma Z} &\simeq 7.6 \times 10^{-29} \, \text{cm}^3\text{s}^{-1}  \frac{y^4}{\mu^2}   \left( \frac{100 \,\text{GeV}}{m_\chi} \right)^2 \,,
\end{align*}
from where it is apparent  the different dependence of the two-to-three and the loop-induced two-to-two processes with $\mu$.

As mentioned above, for (moderately) small mass splittings $m_\psi/m_\chi$ between the fermionic mediator and the scalar dark matter particle, the two-to-three process has the largest cross section, while for $\mu\gtrsim 10$ the one loop processes dominate the annihilation. This is illustrated in Fig.~\ref{fig:vib_vs_loop}, top plots, where we show the annihilation cross sections for the relevant processes as a function of the dark matter mass for $\mu=1.1$ (left plot) and for $\mu=25$ (right plot). It is important to note that the values of the cross section when the loop processes dominate are fairly small. However, this does not imply that the one loop processes can be neglected in this model. As we will show below, gamma-ray lines can dominate the  total gamma-ray energy spectrum, even for smaller values of $\mu$, due to the sharpness of the gamma-ray line signal compared to the internal bremsstrahlung signal and the fairly good energy resolution of present gamma-ray telescopes, $\Delta E\sim 10\%$.

We further investigate the dependence of the cross sections with the parameters of the model in Fig.~\ref{fig:vib_vs_loop}, bottom plots, where we show the cross sections as a function of the degeneracy parameter $\mu$ for two choices of the dark matter mass, $m_\chi=150$ GeV (left plot) and $m_\chi=500$ GeV (right plot). Again, and as apparent from the plots, for choices of parameters where the one loop processes dominate, the resulting cross section is rather small.

\begin{figure*}[t]
\begin{center}
\includegraphics[scale=0.9]{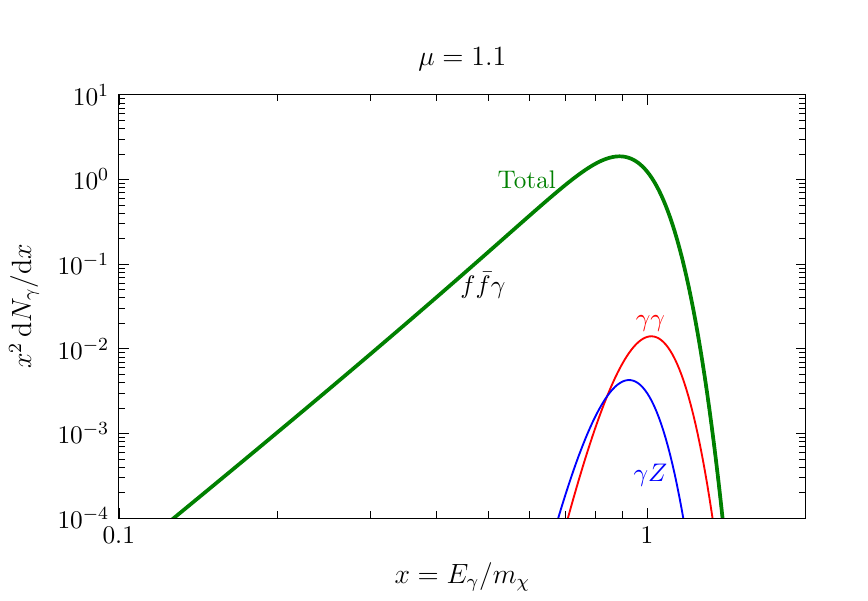}
\includegraphics[scale=0.9]{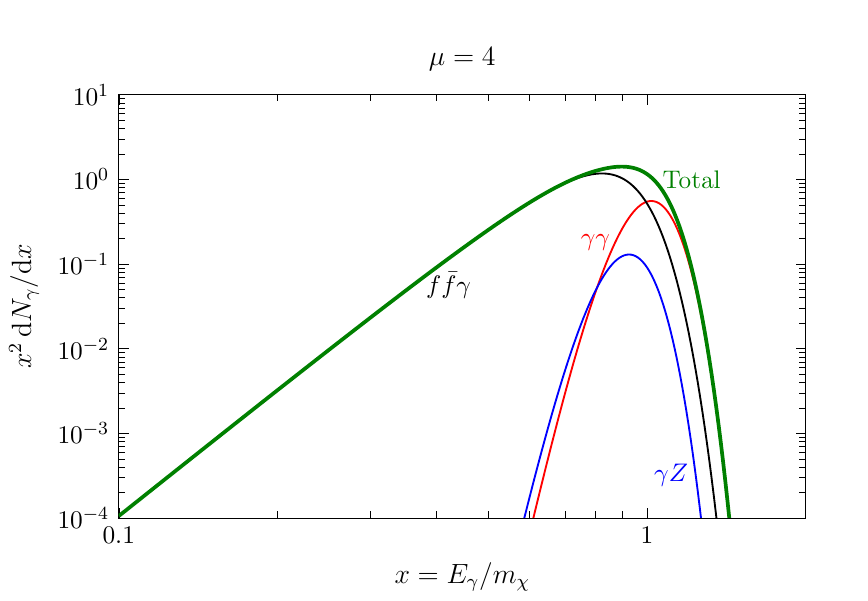}
\includegraphics[scale=0.9]{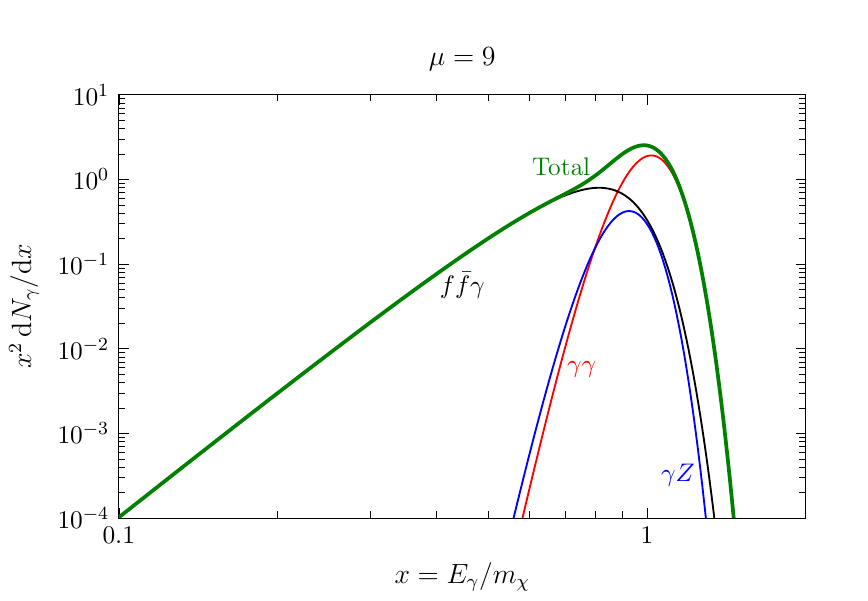}
\includegraphics[scale=0.9]{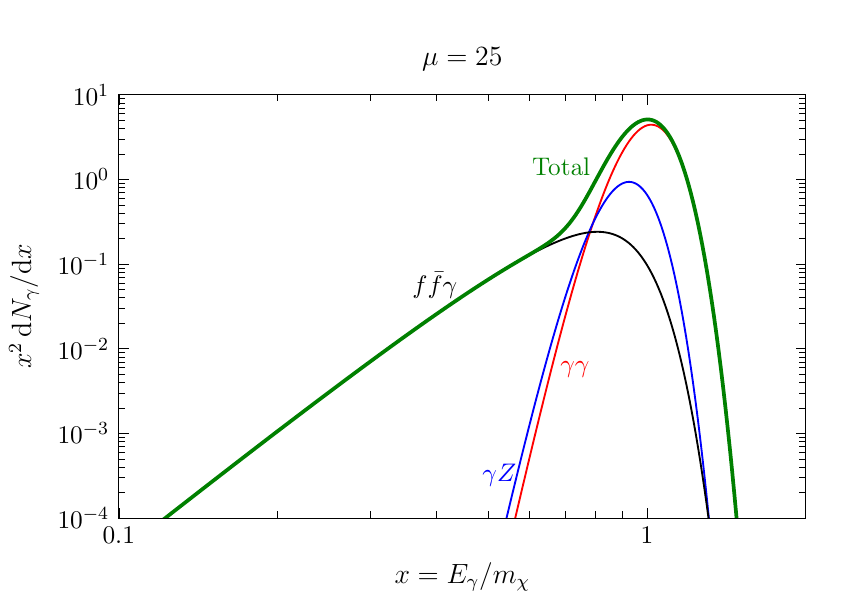}
\caption{Gamma-ray energy spectrum produced in the annihilation into $f\bar f \gamma$ (black line), $\gamma\gamma$ (red line) and $\gamma Z$ (blue line), for $\mu=1.1$ (top left), 4 (top right), 9 (bottom left) and 25 (bottom right), assuming $10\%$ energy resolution.}
\label{fig:spectrum}
\end{center}
\end{figure*}

The total energy spectrum of the channels leading to sharp spectral features reads:
\begin{equation}
\frac{dN_\gamma}{dx}=
\frac{1}{\langle\sigma{v}\rangle}\left[
\frac{d\langle\sigma{v}\rangle_{f\overline{f}\gamma}}{dx}+
2\frac{d\langle\sigma{v}\rangle_{\gamma\gamma}}{dx}+
\frac{d\langle\sigma{v}\rangle_{Z\gamma}}{dx}\right] \,.
\end{equation}
To investigate the relative strength of the gamma-ray line and the internal bremsstrahlung features, we have calculated the gamma-ray energy spectrum for various choices of $\mu$ assuming a Gaussian energy resolution of $10\%$; the result is shown in Fig.~\ref{fig:spectrum} for $\mu=1.1$ (top left), 4 (top right), 9 (bottom left) and 25 (bottom right). It is interesting that the gamma-ray lines give a significant contribution to the spectrum for $\mu\sim 4$ and totally dominate the high energy part of the spectrum when $\mu\gtrsim 9$,  despite the smaller cross section. Therefore, in the search for gamma-ray spectral features the contribution from gamma-ray lines ought not to be neglected. The relative importance of the gamma-ray lines will increase with the next generation of gamma-ray telescopes, such as GAMMA-400~\cite{Galper:2012fp} and DAMPE~\cite{DAMPE}, that aim to an energy resolution of  $\sim 1\%$ at $E_\gamma>10~\mathrm{GeV}$. 

\section{Complementary constraints}
\label{sec:constraints}

The possibility of observing a spectral feature in the gamma-ray sky arising from scalar dark matter annihilations is subject to a series of theoretical and observational constraints. In this section, we discuss each of these constraints individually, while we will study their complementarity in section~\ref{sec:numerics}, along with the results of the searches for gamma-ray spectral features.

\subsection{Perturbativity}
Demanding perturbativity of the model translates into an upper bound on the Yukawa coupling $y$. We will use the common condition $y<4 \pi$ as a conservative upper limit, however, and since the perturbative calculation of the annihilation and scattering rates is performed as an expansion in $\alpha_y\equiv y^2/(4\pi)$, we will also use the condition $y<\sqrt{4\pi}$.

\subsection{Relic density}

The freeze-out of the dark matter particles from the thermal bath, and accordingly the dark matter relic density, is determined by the following annihilation processes: {\it i)} the two-to-two annihilation $\chi\chi\rightarrow f\bar f$ through the Yukawa coupling $y$, {\it  ii)} the higher order processes $\chi \chi \rightarrow f \bar{f} V$ via $t$-channel exchange of the heavy fermion $\psi$ and the one loop annihilations $\chi \chi \rightarrow V V'$ (with $V,V'=\gamma,Z$), {\it iii)} the two-to-two annihilation $\chi \chi \rightarrow h \rightarrow X X'$ via the Higgs portal interactions, where $X$, $X'$ are Standard Model particles. 

We note that the two-to-three annihilation channels $\chi \chi \rightarrow f \bar{f} V$ (and to a lesser extent also the one loop annihilations into a pair of gauge bosons) can contribute significantly to the total cross section at the time of freeze-out, due to the d-wave suppression of the annihilation into $f \bar{f}$. Furthermore, for mass ratios $m_\psi/m_\chi \lesssim 1.2$, the relic density is not set by the annihilation channels listed above, but by other (co-)annihilation processes, the most relevant ones being $\psi \psi \rightarrow f f, \chi \psi \rightarrow f \gamma$ and $\psi \bar{\psi} \rightarrow F \bar{F}$, where $F$ can be any Standard Model fermion. In particular, the latter process is present even for $y = \lambda = 0$. Consequently, for a given mass ratio $m_\psi/m_\chi$, there is a lower bound on the dark matter mass $m_\chi$, below which these annihilation processes are sufficiently strong to suppress the relic density to values smaller than the observed one.

\begin{figure}[t]
\begin{center}
\includegraphics[scale=0.7]{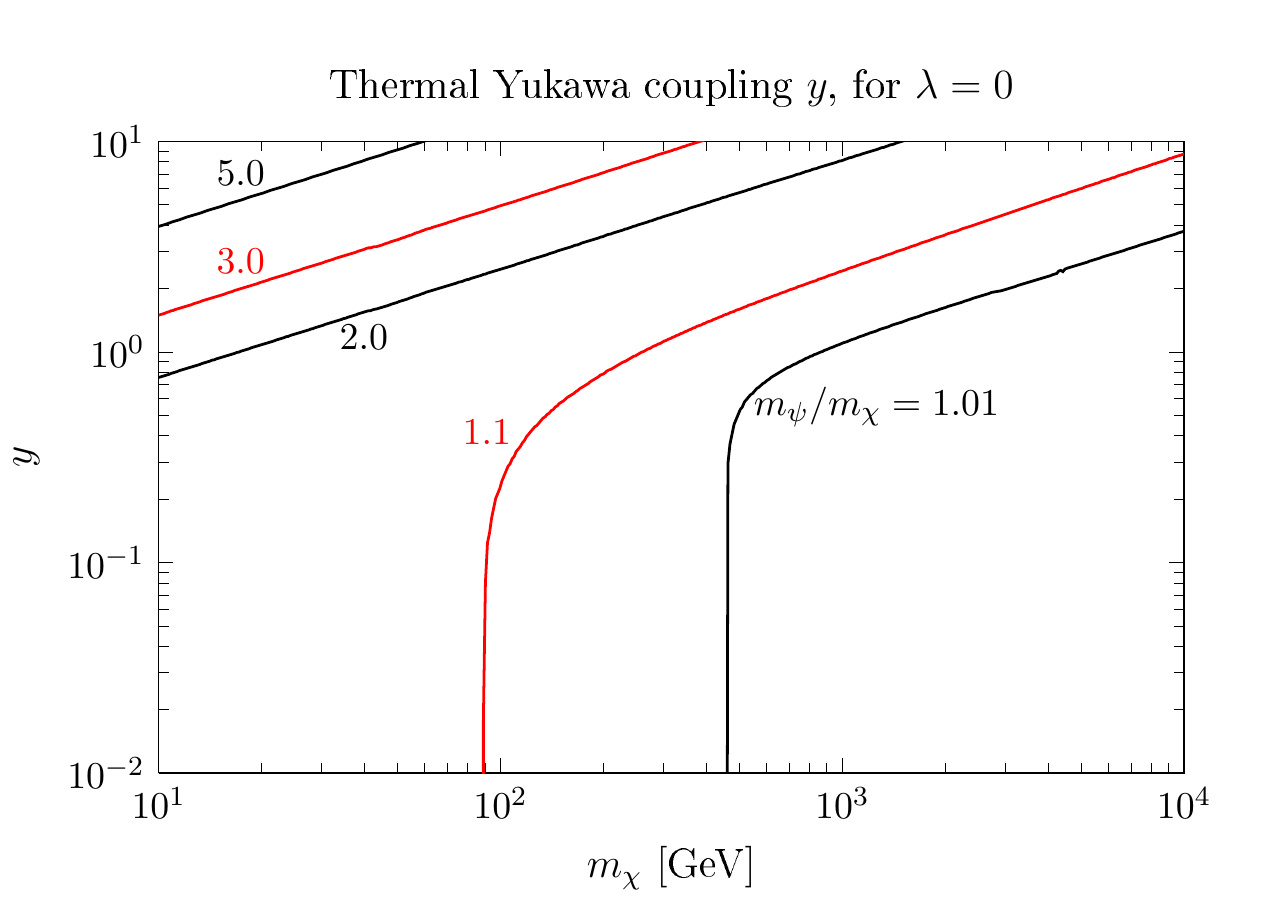}
\caption{Yukawa coupling $y$ leading to the observed dark matter relic density, for $\lambda=0$, and for different values of the mass ratio $m_\psi/m_\chi$.}
\label{fig:yThermal}
\end{center}
\end{figure}

We fully take into account all relevant annihilation and coannihilation channels by adapting the micrOMEGAs code~\cite{Belanger:2013oya} for the solution of the Boltzmann equation, taking $\Omega_{\rm DM} h^2 \simeq 0.12$ as measured by Planck~\cite{Ade:2013zuv}. In this way, also the increase of the annihilation cross section close to the Higgs resonance is included, which suppresses the relic density for $m_\chi \simeq m_h/2$ and values slightly below. For $y = 0$, our results agree reasonably well with~\cite{Cline:2013gha}. In Fig.~\ref{fig:yThermal} we show the Yukawa coupling $y$ leading to the observed dark matter relic density, for $\lambda=0$ and different choices of the mass ratio $m_\psi/m_\chi$. Setting the scalar coupling $\lambda$ to non-zero values leads to a decrease of $y_\text{thermal}$ with respect to the values shown in Fig.~\ref{fig:yThermal}. This will be further discussed  in section~\ref{sec:numerics}.\newpage

\subsection{Direct Detection}

The dark matter particle $\chi$ can scatter off nuclei via $t$-channel Higgs exchange with quarks\footnote{In principle, also scattering processes involving the Yukawa coupling to the SM lepton $f$ are possible; however, the corresponding scattering cross section is heavily suppressed as the first non-zero contribution arises from two-loop diagrams~\cite{Kopp:2009et}.}. The corresponding spin independent scattering cross section with nucleons is given by
\begin{equation}
\sigma_{\text{SI}}=\frac{\lambda^2 f_N^2}{4\pi} \frac{\mu_N^2 m_N^2}{m_h^4 m_\chi^2} \,,
\end{equation}
where $\mu_N=m_N m_\chi / \left(m_N + m_\chi\right)$ is the DM-nucleon reduced mass, and $f_N=\sum_q f_q$ is the Higgs-nucleon coupling. The latter is subject to significant nuclear uncertainties, in particular it is sensitive to the strangeness content $f_s$ of protons and neutrons. In the following we use the value $f_N=0.345$, as determined in the recent study~\cite{Cline:2013gha}. Furthermore, we fix the Higgs mass to be $m_h=125$ GeV. 

\begin{figure}[t]
\begin{center}
\includegraphics[scale=0.7]{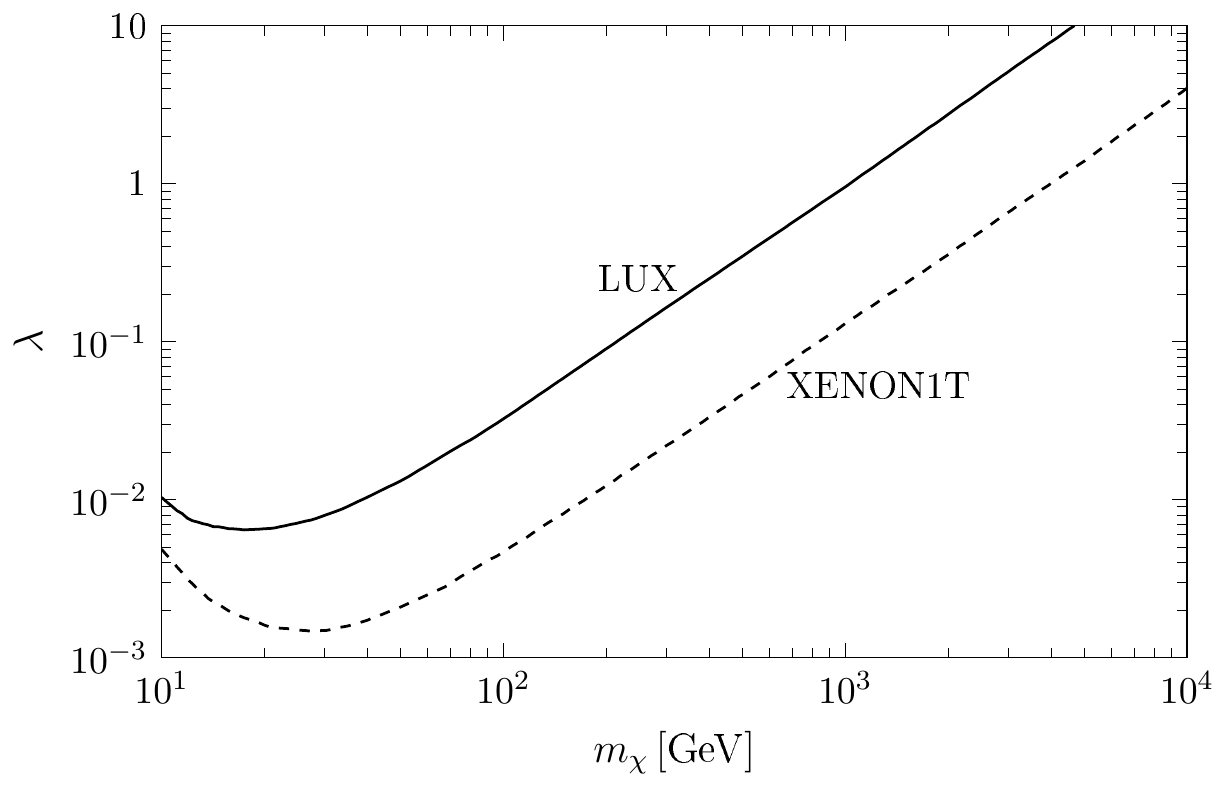}
\caption{Upper limit on the scalar coupling $\lambda$ as a function of the dark matter mass derived from the LUX data (solid line) as well as the prospected sensitivity of the XENON1T experiment (dashed line).}
\label{fig:LamdaConstraintsDD}
\end{center}
\end{figure}

The present upper limit from the LUX experiment~\cite{Akerib:2013tjd} on the spin independent cross section can be used to set a limit on the scalar coupling $\lambda$ as a function of the dark matter mass $m_\chi$, independently of the values of  $y$ and $m_\psi$. This limit will be improved in the future by the  XENON1T experiment~\cite{Aprile:2012zx}, which will have a sensitivity to $\sigma_{\text{SI}}$ which is approximately a factor 100 better than the current XENON100 experiment~\cite{Aprile:2012nq}. The present LUX limit, as well as an estimate on the projected sensitivity to $\lambda$ of the XENON1T experiment, are shown in Fig.~\ref{fig:LamdaConstraintsDD} as a solid and a dashed line, respectively.

\subsection{Indirect detection}
In addition to the sharp gamma-ray spectral features, the scalar dark matter model  we consider in this paper leads to other potentially observable signals in indirect detection experiments, such as exotic contributions to the antimatter fluxes or the continuum gamma-ray flux. We will consider first the scenario where the Higgs portal coupling is set to zero and will discuss later on the impact of a sizable $\lambda$  on the fluxes.

The annihilation channel $f \bar{f} \gamma$ is necessarily accompanied by the process $f \bar{f} Z$, leading to the production of antiprotons via the decay and hadronization of the $Z$ boson; this has been studied in detail in the context of Majorana dark matter~\cite{Kachelriess:2009zy,Ciafaloni:2011sa,Garny:2011cj,Garny:2011ii}. In the case of the scalar dark matter model at hand, we find for the differential cross section the expression
%\begin{widetext}
\begin{align}
&\frac{\text{d}\sigma{v}_{f\overline{f}Z}}{\text{d}x \,\text{d}z}=
\frac{y^4\alpha_\mathrm{em}\tan^2\theta_W}
{\pi^2m_\chi^2\left(1-\mu-2z\right)^2\left(3+\mu-2x-2z\right)^2} \times \nonumber\\
&\left\{(1-x)\left[x^2-2x(1-z)+2(1-z)^2\right]+\frac{\xi}{4}(x^2-2x+2)\right\},
\end{align}
%\end{widetext}
where $x=E_Z/m_\chi$, $z=E_f/m_\chi$ and $\xi=m_Z^2/m_\chi^2$. Note that, while the differential cross section for $f \bar{f} \gamma$ in the case of scalar DM is simply a factor of 8 larger than for Majorana DM, this expression for $f \bar{f} Z$ has a different functional dependence on $x$ and $y$ compared to the corresponding process for Majorana DM (see e.g.~\cite{Garny:2011ii} for the latter). The one loop annihilations into $\gamma Z$ and $ZZ$ also contribute to the antiproton flux, however, these channels are only relevant when $m_\psi/m_\chi \gtrsim 3$. In this regime, the resulting antiproton flux is too small to be observed, therefore, the one loop annihilation channels will be neglected in our analysis.

\begin{figure*}[t]
\begin{center}
\includegraphics[scale=0.45]{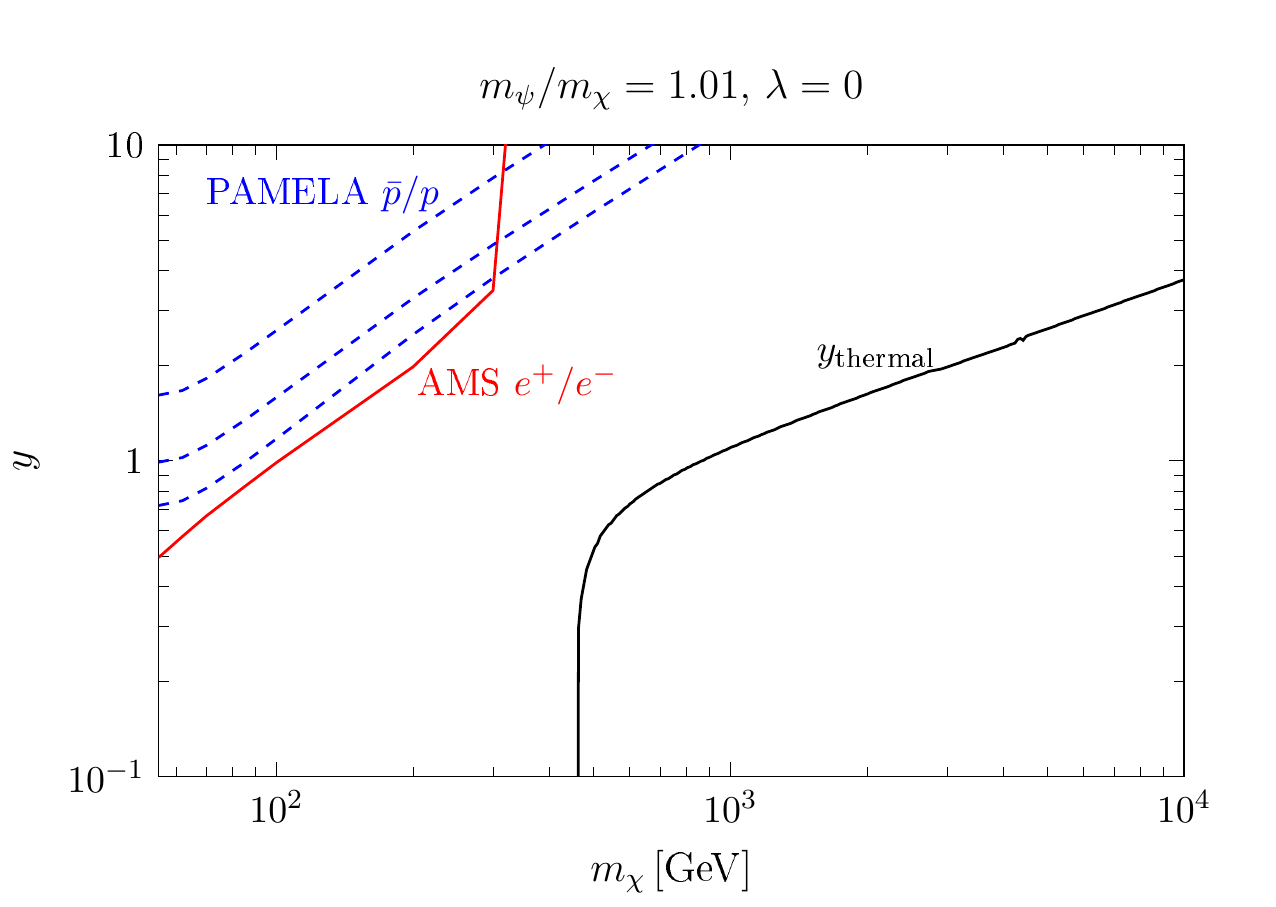}
\includegraphics[scale=0.45]{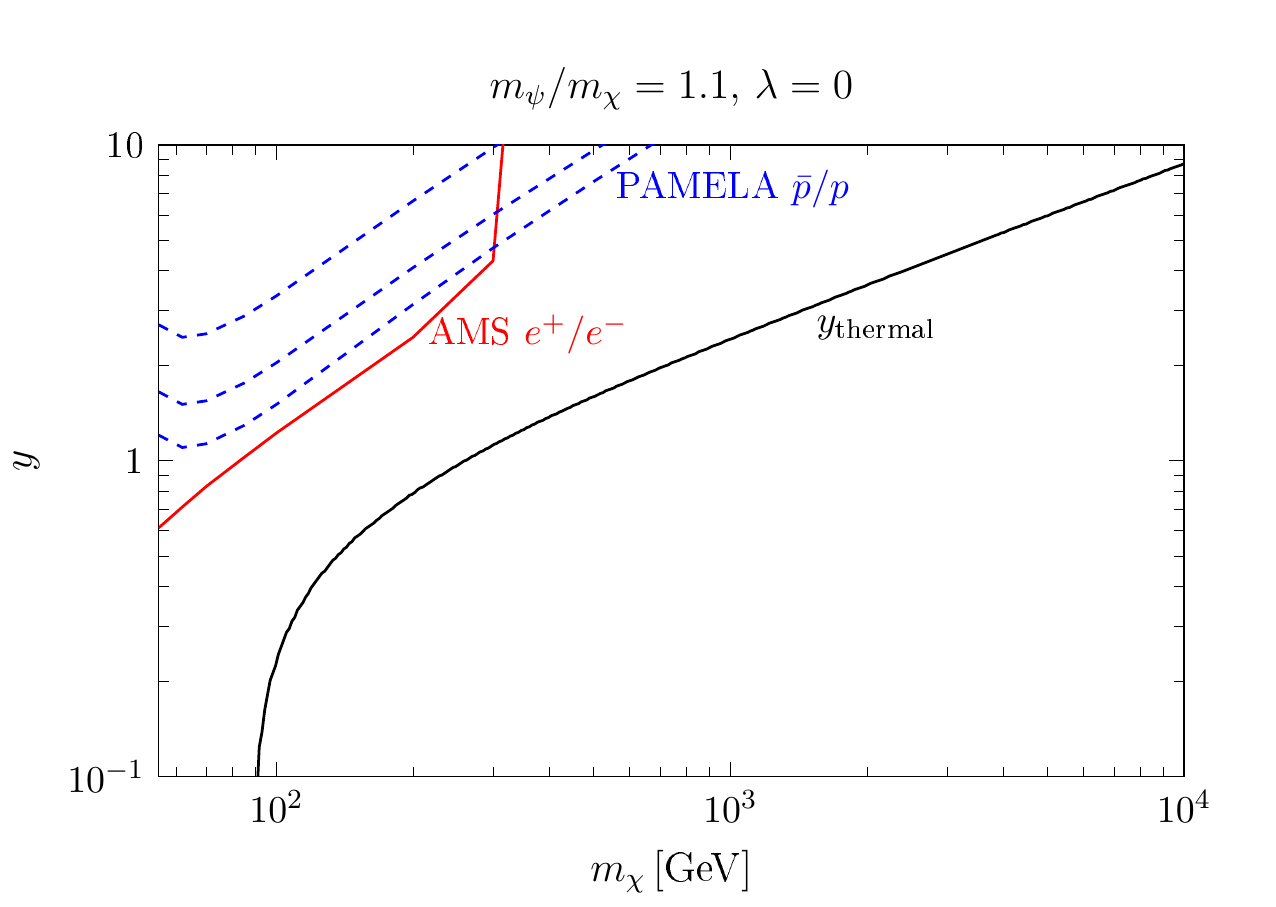}
\includegraphics[scale=0.45]{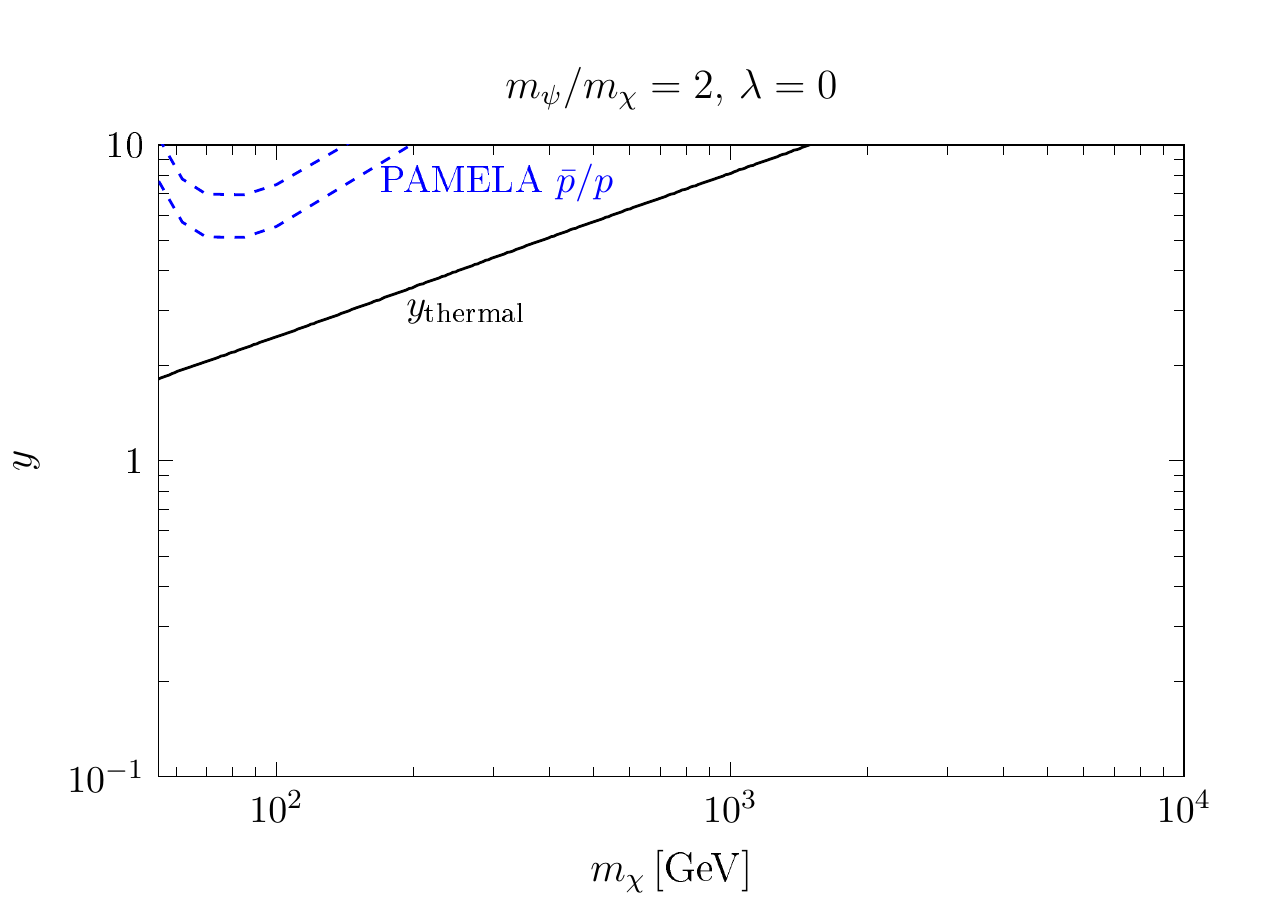}
\caption{95 $\%$ C.L. upper limit on the Yukawa coupling $y$ as a function of the dark matter mass for $m_\psi/m_\chi=1.01$ (left plot), 1.1 (middle plot) and 2 (right plot), derived from the non-observation in the PAMELA data on $\bar p/p$ of a significant excess of antiprotons produced in the annihilation $\chi\chi\rightarrow f\bar f Z$, with $f=e,\,\mu,\,\tau$ (dashed blue lines, corresponding from top to bottom to the MIN, MED and MAX propagation parameters), as well as from the non-observation in the AMS-02 data on the positron fraction of the spectral feature generated in the annihilation $\chi\chi\rightarrow e^- e^+  \gamma$ (red solid line). The value of the Yukawa coupling required to generate the observed relic abundance via thermal production is shown, for comparison, as a black solid line. In all plots it is assumed $\lambda=0$.}
\label{fig:antimatter_constraints}
\end{center}
\end{figure*}

We use CalcHEP \cite{Pukhov:1999gg,Pukhov:2004ca} interfaced with PYTHIA \cite{Sjostrand:2007gs} to produce the corresponding injection spectrum of antiprotons. The propagation to the Earth is implemented by using the standard two-zone diffusion model, neglecting energy losses and reacceleration; hereby, we use the same setup as in~\cite{Garny:2011ii} to which we refer the reader for details. We employ the MIN, MED and MAX propagation parameters from~\cite{Donato:2003xg} in order to bracket the uncertainty arising from the different possible parameters of the propagation model; furthermore, the analysis is done assuming a NFW halo profile, although this choice does not affect the limits significantly. Finally, solar modulation is included by means of the force-field approximation~\cite{1967ApJ149L115G} with $\Phi_{F} = 500$ MeV. We then compare the sum of the propagated antiproton spectrum and the expected spallation background, which we take from \cite{Bringmann:2006im}, with the PAMELA $\bar{p}/p$ data \cite{Adriani:2010rc}. We perform a $\chi^2$-test at $95 \%$ C.L. in order to obtain the corresponding constraints on the Yukawa coupling $y$. The resulting limit is shown in Fig~\ref{fig:antimatter_constraints} as a function of the dark matter mass for $m_\chi/m_\chi=1.01$ (left plot), 1.1 (middle plot) and 2 (right plot). We also show for comparison the values of the Yukawa coupling $y$ leading to the observed relic abundance via thermal freeze-out for any values of $m_\chi$ and $m_\psi$. As apparent from the plots, present experiments are not sensitive enough to detect the antiprotons produced in the annihilations of scalar dark matter particles.

Besides antiprotons, also electrons and positrons can be produced in the annihilation processes. However, in contrast to the case of antiprotons, the observed positron fraction $e^+/\left(e^- + e^+ \right)$ is not compatible with the expectation from purely secondary positron production by spallation of cosmic rays. While the reason for the unexpected rise of the positron fraction as observed by PAMELA~\cite{Adriani:2008zr}, Fermi-LAT~\cite{FermiLAT:2011ab} and AMS-02~\cite{Aguilar:2013qda} remains unclear, the extremely precise data from AMS-02 can be used for setting strong limits on the annihilation cross section of dark matter into leptonic final states, notably when $m_{\text{DM}} \lesssim 300$ GeV~\cite{Bergstrom:2013jra,Ibarra:2013zia}.

In our scenario, for $f=e^-$ and $m_\psi/m_\chi \simeq 1$, the injection spectrum of the positrons arising in the annihilation process $\chi \chi \rightarrow e^- e^+ \gamma$ exhibits a pronounced spectral feature towards the kinematical end point, which makes it separable from the smooth background, even after taking into account propagation effects. In the following, we use the limits on $\langle \sigma v \rangle_{e^- e^+ \gamma}$ derived in~\cite{Bergstrom:2013jra}, and convert them into limits on the Yukawa coupling $y$, neglecting the subdominant process $\chi \chi \rightarrow e^- e^+ Z$. It is important to note that the limit in~\cite{Bergstrom:2013jra} was derived under the assumption $m_\psi \equiv m_\chi$, leading to the above-mentioned sharp spectral feature in the positron spectrum. Up to mass ratios $m_\psi/m_\chi \lesssim 1.2$, the spectrum of the positrons does not change significantly and our procedure of obtaining the corresponding limits on the Yukawa coupling is applicable; for larger values of $m_\psi/m_\chi$, the positron spectrum becomes broader, making it harder to distinguish from the background. The limits for $\mu=1.01$ and 1.1 are shown, respectively, in the left and middle plots of Fig.~\ref{fig:antimatter_constraints} and, as for the antiprotons, are well above the value required by the freeze-out mechanism.

\begin{figure}[t]
\begin{center}
\includegraphics[scale=0.85]{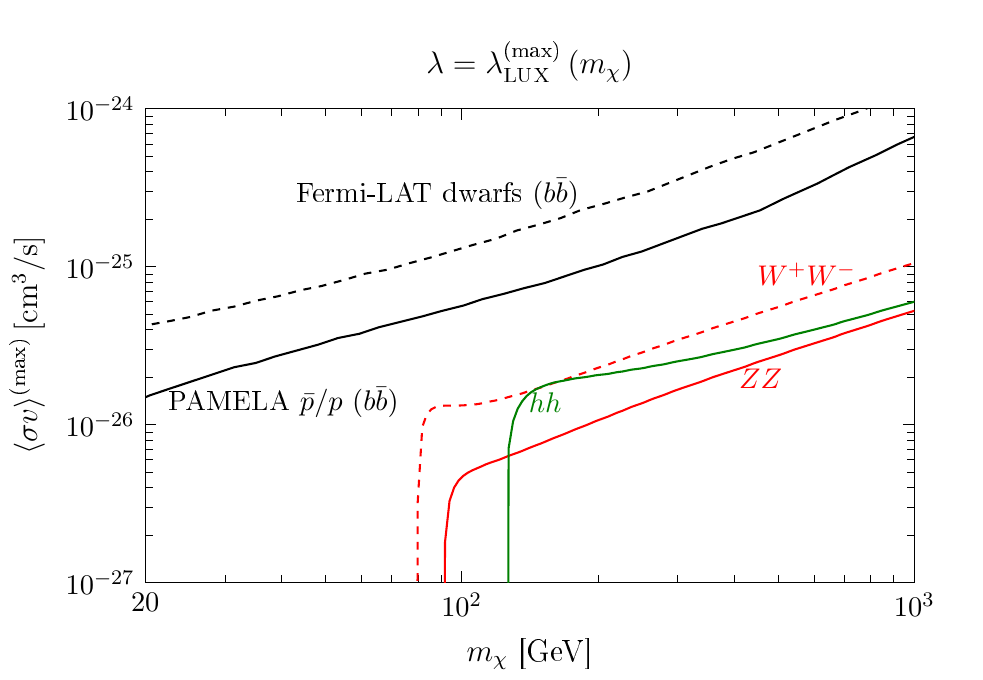}
\caption{Maximally allowed annihilation cross sections for the Higgs-mediated processes, imposing the upper limit on $\lambda$ from the LUX experiment, together with bounds from PAMELA and Fermi-LAT.}
\label{fig:IndirectConstraintsMaxLambda}
\end{center}
\end{figure}

For $\lambda > 0$, new annihilation channels mediated by the Standard Model Higgs open up, the most relevant ones being $\chi \chi \rightarrow W^+ W^-$, $\chi \chi \rightarrow ZZ$ and $\chi \chi \rightarrow hh$. In the limit $v\rightarrow 0$, the annihilation cross sections read

\begin{align}
\langle \sigma v \rangle_{ZZ} &= \frac{\lambda^2 \left( 4 m_\chi^4 - 4 m_\chi^2 m_Z^2 + 3 m_Z^4 \right) \sqrt{m_\chi^2-m_Z^2}}{16 \pi m_\chi^3 \left( m_h^2-4 m_\chi^2\right)^2 } \,, \nonumber\\
\langle \sigma v \rangle_{W^+ W^-} &= \frac{\lambda^2 \left( 4 m_\chi^4 - 4 m_\chi^2 m_W^2 + 3 m_W^4 \right) \sqrt{m_\chi^2-m_W^2}}{8 \pi m_\chi^3 \left( m_h^2-4 m_\chi^2\right)^2 } \,, \nonumber\\
\langle \sigma v \rangle_{hh} &= \frac{\sqrt{m_\chi^2 -m_h^2}}{16 \pi m_\chi^3 \left( 8 m_\chi^4 -6m_\chi^2 m_h^2+m_h^4\right)^2} \times \nonumber\\
&\hspace{-0.5cm}\lambda^2 \left( 4 m_\chi^4 +4 \lambda m_\chi^2 v_{\text{EW}}^2-m_h^2 \left[ m_h^2+\lambda v_{\text{EW}}^2 \right] \right)^2 \,,
\end{align}
with $v_{\text{EW}} = 246$ GeV being the vacuum expectation value of the Standard Model Higgs. These annihilation channels contribute to the antiproton flux, and also lead to a featureless gamma-ray spectrum. In Fig.~\ref{fig:IndirectConstraintsMaxLambda}, we show the annihilation cross sections into these final states, fixing $\lambda$ to be the upper limit deduced from the LUX experiment (as shown in  Fig.~\ref{fig:LamdaConstraintsDD}). When comparing to the upper bounds derived in~\cite{Fornengo:2013xda} from the PAMELA data on the antiproton-to-proton fraction, as well as the latest Fermi-LAT limits obtained by searches for gamma-ray emission from dwarf galaxies~\cite{Ackermann:2013yva}, we find that these limits from indirect detection are always superseded by the upper limit on $\lambda$ from the LUX experiment. Note that for simplicity, we only show the PAMELA and Fermi-LAT bounds for the annihilation channel $\chi \chi \rightarrow b \bar{b}$; the constraints for the relevant annihilation channels listed above are all comparable~\cite{Fornengo:2013xda,Ackermann:2013yva}, leading to the same conclusion.

\begin{figure*}[t]
\begin{center}
\includegraphics[scale=0.65]{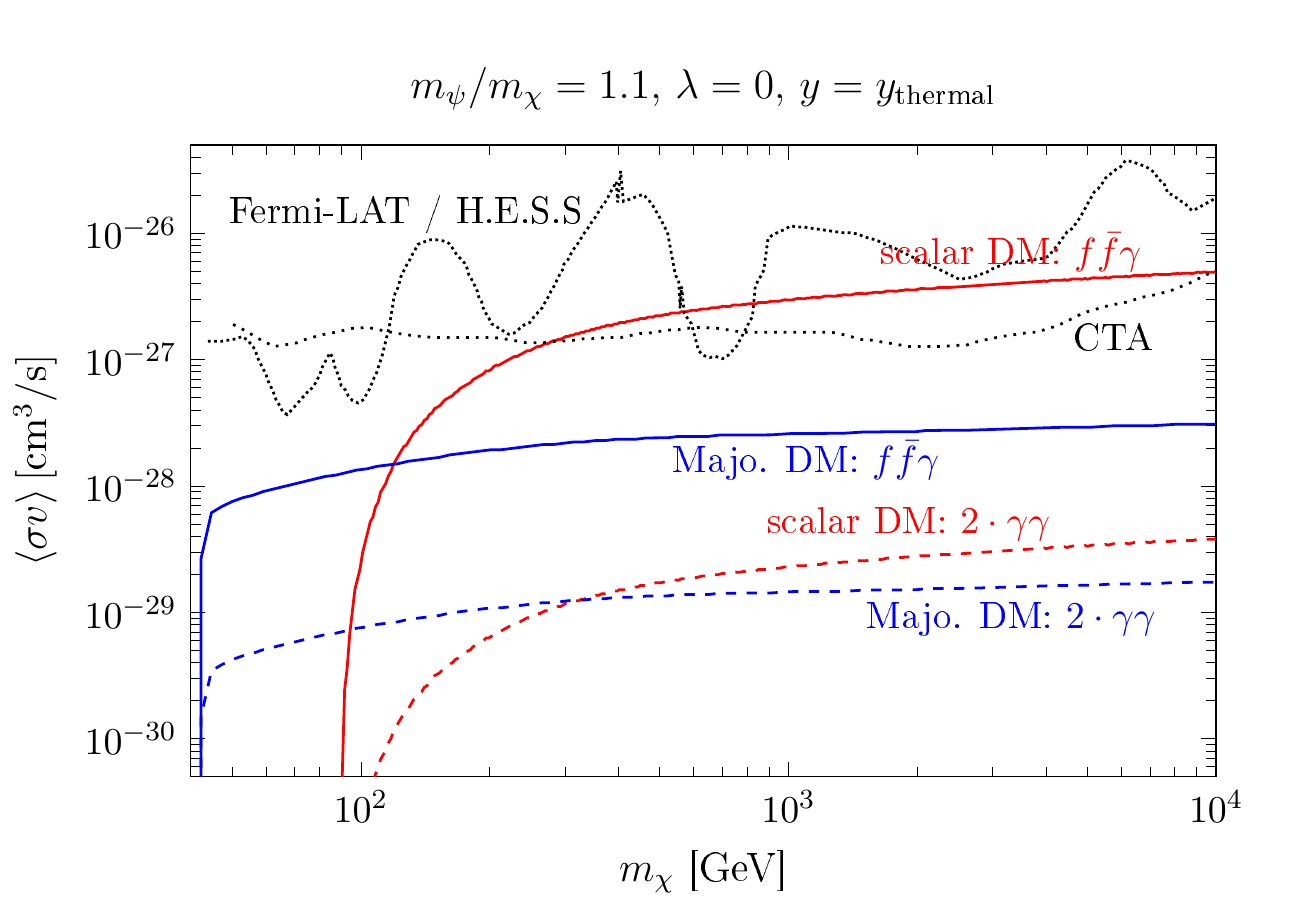}
\includegraphics[scale=0.65]{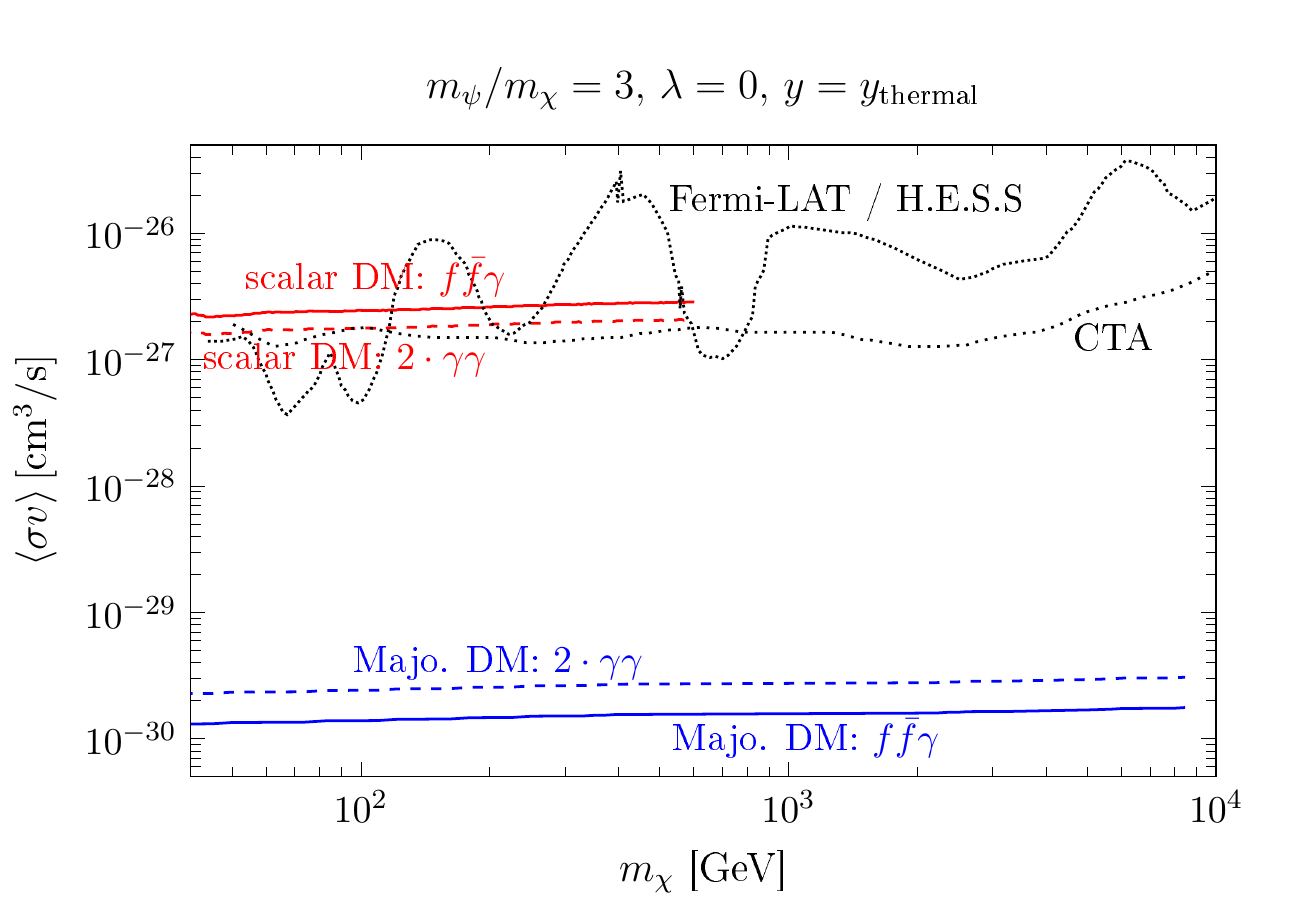}
\caption{Annihilation cross sections $\langle \sigma v \rangle$ into final states leading to sharp gamma-ray spectral features, for scalar dark matter (red lines) and Majorana dark matter (blue lines). In both cases, for each value of the dark matter mass $m_\chi$, the Yukawa coupling $y$ is fixed to the value leading to the observed relic density. All lines are only shown for cases corresponding to $y < 4 \pi$. The black dotted lines are upper limits from Fermi-LAT, H.E.S.S. and CTA (see main text).}
\label{fig:sigmav_thermal}
\end{center}
\end{figure*}

\subsection{Collider Constraints}
The scalar dark matter model can be probed at colliders,
most notably at the LHC. A full analysis of the collider constraints is
beyond the scope of this work, and we focus on the Drell-Yan production
of a pair of heavy fermions $\psi\overline{\psi}$. These particles in
turn both decay into the dark matter particle $\chi$ and the Standard
Model lepton $f$ (or $\bar{f}$), leading to a charged lepton pair plus
missing energy. This signal is similar to the slepton search at
LHC, however the production cross
section for the charged fermion is about one order of magnitude larger
than for a charged scalar~\cite{Bai:2014osa}, leading to tighter
constraints. 

We calculate the production cross section $pp\to\psi\overline{\psi}$
using CalcHEP~\cite{Pukhov:1999gg,Pukhov:2004ca}, and compare it with
the upper bound of the production cross section along with the analysis in
ref.~\cite{Chang:2014tea}. The computed production cross section depends only the heavy fermion mass $m_\psi$. The experimental data is taken from ATLAS with
the integrated luminosity of $20.3~\mathrm{fb^{-1}}$ and
$\sqrt{s}=8~\mathrm{TeV}$~\cite{TheATLAScollaboration:2013hha}. 
Additionally, we estimate a bound from the LEP experiment, based on a 
search for the right-handed selectron in the minimal supersymmetric Standard
Model, with data collected in the energy range $\sqrt{s}=183-208~\mathrm{GeV}$~\cite{LEP_bound}. 
This bound does not exactly apply to the model discussed in this work, and hence only serves as an estimation.

\section{Constraints for thermally produced scalar dark matter}
\label{sec:numerics}

We present in this section the constraints on the parameter space of the scalar dark matter model from the negative searches for gamma-ray spectral features, assuming that the dark matter density was generated via thermal freeze-out, as well as the complementarity of these constraints with those discussed in section~\ref{sec:constraints} from direct detection and collider experiments. Let us discuss separately the scenario where the scalar coupling $\lambda$ is small and where it is sizable.

\subsection{Small scalar coupling $\lambda$}

\begin{figure*}[t]
\begin{center}
\includegraphics[scale=0.9]{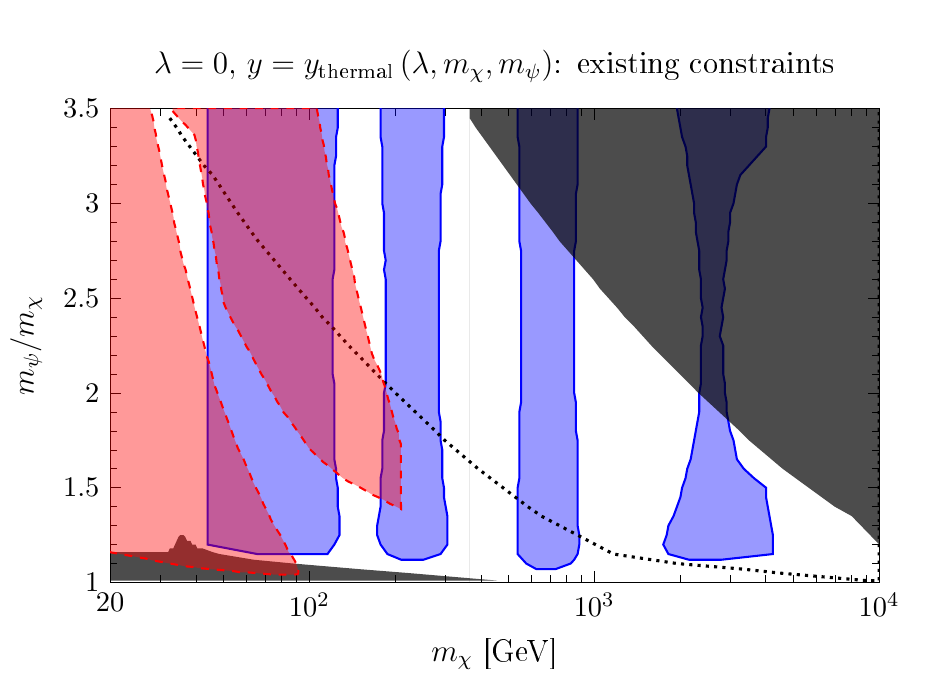}
\includegraphics[scale=0.9]{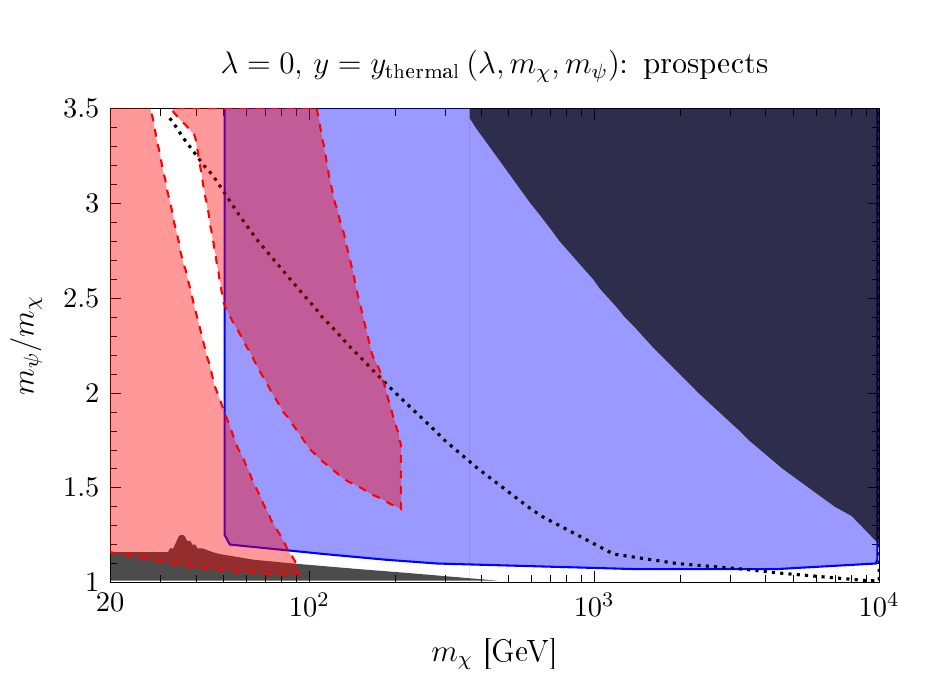}
\caption{Left plot: constraints from Fermi-LAT and H.E.S.S (blue shaded regions surrounded by solid lines), from LEP and LHC (red shaded regions enclosed by dashed lines), setting at each point of the parameter space the Yukawa coupling in order to reproduce the observed dark matter abundance via thermal production. We assume in the plots $\lambda=0$. The dark gray shaded regions extending to the right upper corner correspond to scenarios violating the conservative perturbativity requirement $y > 4 \pi $, while the dotted black line corresponds to scenarios with $y = \sqrt{4 \pi}$. The gray shaded regions in the lower left corner show the values of the parameters for which the dark matter abundance is below the observed value due to coannihilations. Right plot: same as left plot, but here the blue shaded region (enclosed by the solid line) is the prospected reach of CTA.}
\label{fig:lambda_zero}
\end{center}
\end{figure*}

Let us discuss first the case where the coupling $\lambda$ between the dark matter particle $\chi$ and the Standard Model Higgs doublet is very small, concretely $\lambda\lesssim 10^{-3}$. In this limit, and as argued in section~\ref{sec:constraints}, the Higgs portal interaction is too suppressed to lead to observable signatures in the direct detection experiments LUX and XENON1T or in the indirect detection experiments AMS-02 or PAMELA. In this case, the requirement of reproducing the observed dark matter abundance via thermal production fixes the Yukawa coupling of the model as a function of $m_\chi$ and $m_\psi$. 

We show  in Fig.~\ref{fig:sigmav_thermal} the annihilation cross sections into $f \bar{f} \gamma$ and $\gamma \gamma$ for $\lambda=0$ and for $m_\psi/m_\chi=1.1$ (left plot) and 3 (right plot) as a function of $m_\chi$ in the scalar dark matter scenario. These cross sections can be compared to the searches for sharp spectral features in the gamma-ray sky performed by the Fermi-LAT~\cite{Ackermann:2013uma} and H.E.S.S.~\cite{Abramowski:2013ax} collaborations. Upper limits on the combined annihilation cross section $\langle \sigma v \rangle_{f \bar{f} \gamma} + 2 \langle \sigma v \rangle_{\gamma \gamma}$ were derived in~\cite{Garny:2013ama}, employing the spectrum of gamma-rays originating from a region close to the Galactic Center, for a model of Majorana dark matter which couples to a heavy scalar and a right-handed Standard Model fermion $f_R$ via a Yukawa coupling. Since the normalized energy spectrum of gamma-rays from the two-to-three processes is identical for the cases of scalar and Majorana dark matter, we adopt those limits for our analysis. These constraints are shown in Fig.~\ref{fig:sigmav_thermal}, together with the prospected sensitivity of CTA~\cite{Bernlohr:2012we}, which we also take from~\cite{Garny:2013ama}.\footnote{Strictly, the limits for the Majorana case can only be applied to the scalar case when the fermionic mediator and the dark matter particle are close in mass, $m_\psi/m_\chi \lesssim 2-3$, namely when the two-to-three annihilations into $f \bar{f} \gamma$ dominate over the one loop annihilations into two photons. For larger mass ratios, $m_\psi/m_\chi \gtrsim 3$, the relative importance of the two-to-three and one loop annihilations is different for Majorana and for scalar dark matter and the limits for the former cannot be straightforwardly applied for the latter. Due to the mild dependence of the upper limits on $\langle \sigma v \rangle$ on the mass ratio~\cite{Garny:2013ama}, we estimate that for $m_\psi/m_\chi>3$ the actual upper limits for the scalar dark matter model deviate from these results by at most a factor of two.} It follows from  Fig.~\ref{fig:sigmav_thermal} that the Fermi-LAT and H.E.S.S. observations are currently probing regions of the parameter space where the scalar dark matter particle is thermally produced, as first mentioned in~\cite{Toma:2013bka, Giacchino:2013bta}. We also show for comparison the corresponding thermal annihilation cross sections for the analogous model involving a Majorana dark matter candidate, as studied in e.g.~\cite{Garny:2013ama}, and which is not expected to produce any observable signal in current- or next-generation experiments searching for gamma-ray spectral features, unless a boost factor of the Galactic gamma-ray signal  is invoked~\cite{Garny:2013ama}. 

The complementarity of the various search strategies is illustrated in Fig.~\ref{fig:lambda_zero}, where we show the excluded regions in the parameter space of the scalar dark matter model, spanned by the mass ratio $m_\psi/m_\chi$ and dark matter mass $m_\chi$ (assuming thermal production). In the left panel, the blue shaded regions enclosed by the solid lines are excluded by Fermi-LAT or H.E.S.S., while the red shaded regions surrounded by dashed lines are excluded by collider searches. The dark gray shaded regions in the right upper part of each plot are ruled out by the conservative perturbativity requirement $y< 4\pi$ while the dotted black line shows the points with $y=\sqrt{4\pi}$. Lastly, the gray shaded regions in the lower left corners of the parameter space correspond to choices of parameters where the measured dark matter abundance cannot be generated via thermal freeze-out, due to very efficient coannihilations. It follows from the figures  that the parameter space for thermally produced scalar dark matter particles is considerably constrained by the requirement of perturbativity of the theory, by collider searches and by the searches for sharp gamma-ray spectral features. In the right panel of Fig.~\ref{fig:lambda_zero}, we show the reach of CTA on the scalar dark matter model, assuming again thermal production. Remarkably, CTA has good prospects to probe practically the whole parameter space, in particular values of the dark matter mass which are inaccessible to present and projected collider searches.

\subsection{Sizable scalar coupling $\lambda$}

If the coupling $\lambda$ is sizable, the direct detection signals can be significantly enhanced, as discussed in section~\ref{sec:constraints}. Moreover, the Higgs portal coupling opens new additional annihilation channels at the time of freeze-out. Therefore, and in order to reproduce to observed dark matter abundance, the Yukawa coupling $y$ must be smaller compared to the case $\lambda=0$. Accordingly, the intensity of the gamma-ray spectral features is expected to be smaller in this case than when $\lambda=0$.

\begin{figure*}[t]
\begin{center}
\includegraphics[scale=0.9]{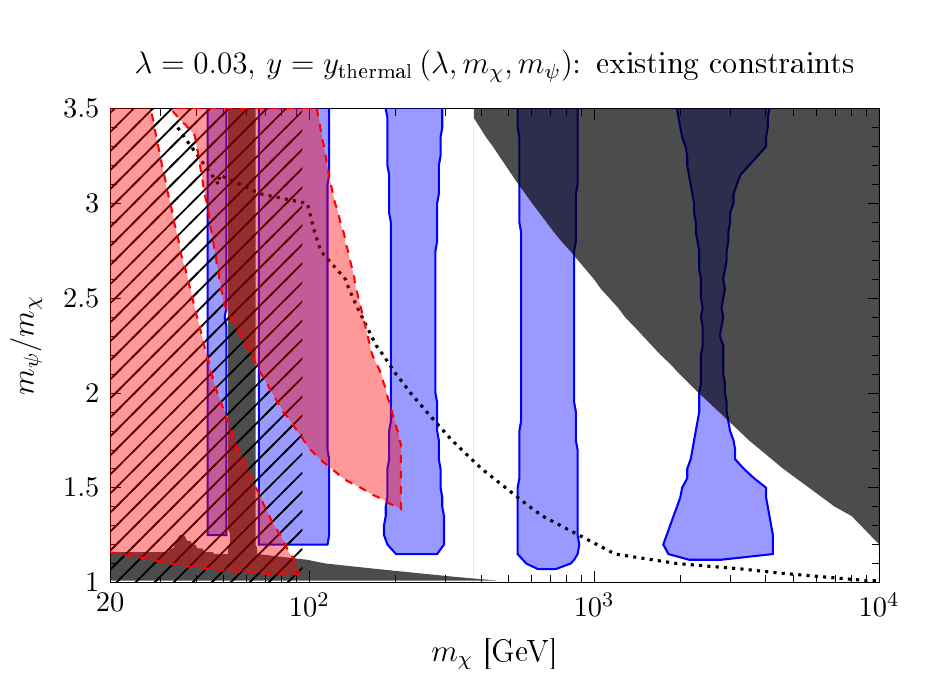}
\includegraphics[scale=0.9]{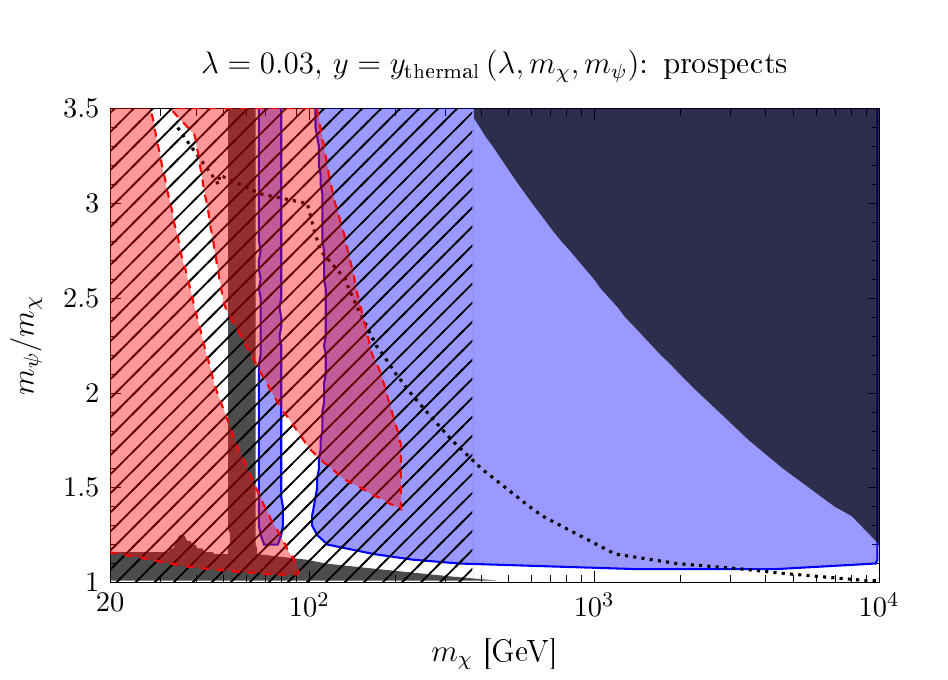}\\
\includegraphics[scale=0.9]{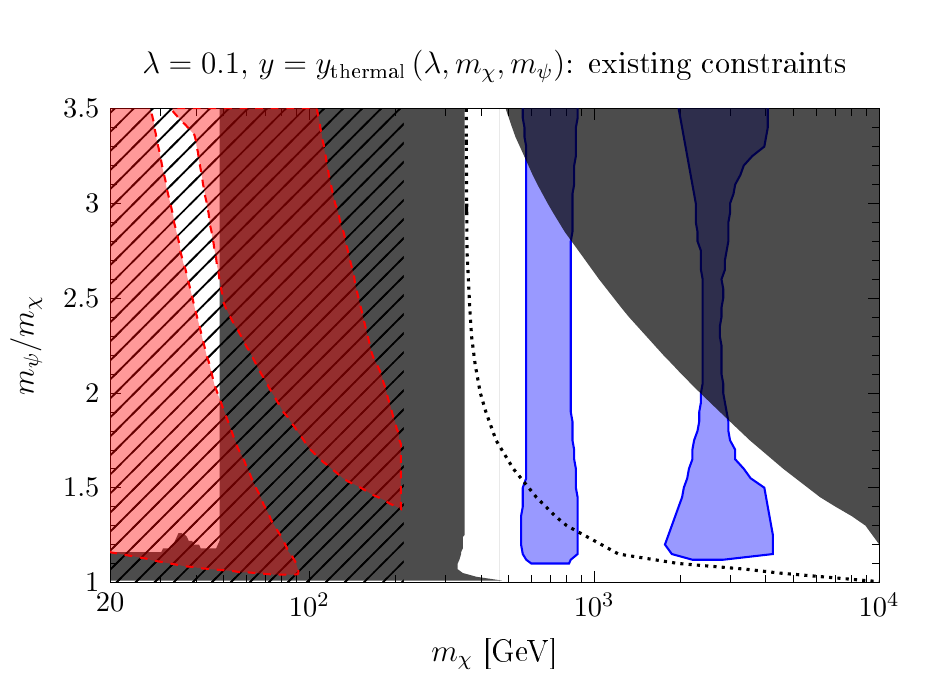}
\includegraphics[scale=0.9]{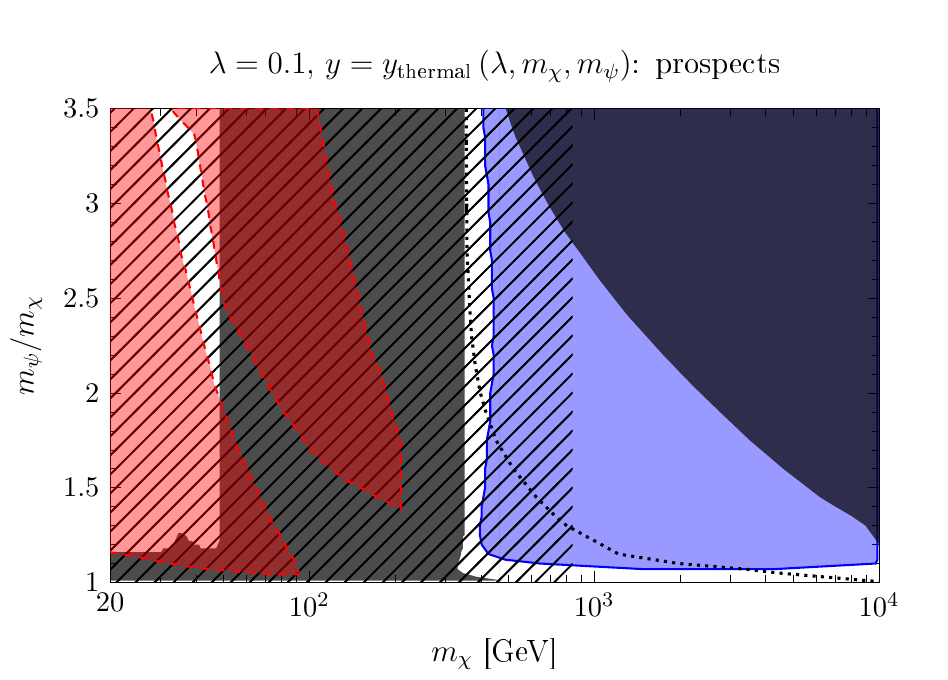}
\caption{Same as Fig.~\ref{fig:lambda_zero}, but adopting a non-vanishing value for the Higgs portal coupling, concretely $\lambda=0.03$ (upper panels) and 0.1 (lower panels). The hatched regions in the left plot are excluded by LUX, while in the right plot are within the reach of XENON1T.}
\label{fig:lambda_finite}
\end{center}
\end{figure*}

We show in Fig.~\ref{fig:lambda_finite} the parameter space of the scalar dark matter model, spanned by $m_\psi/m_\chi$ and $m_\chi$, for the value of the Yukawa coupling $y$ leading to the observed dark matter abundance for a fixed value of the coupling $\lambda=0.03$ (upper panels) or 0.1 (lower panels). Compared to Fig.~\ref{fig:lambda_zero}, which assumed $\lambda=0$, we also include in the plots the choices of parameters excluded by the LUX experiment (left panels) and the projected sensitivity of the XENON1T experiment (right panels), shown as a hatched region. Moreover, depending on the value of $\lambda$, there is a range of dark matter masses $m_\chi$ for which the Higgs-mediated annihilation channels are sufficiently strong to suppress the relic density below its observed value, independently of the Yukawa coupling $y$. This is indicated in Fig.~\ref{fig:lambda_finite} by the (nearly) vertical gray shaded strips. In particular, it is impossible to obtain the observed relic density for dark matter masses around $m_\chi \simeq m_h/2 \simeq 63$ GeV, even for very small values of $\lambda$, due to the resonantly enhanced annihilation processes mediated by the exchange of a Higgs particle in the s-channel.

It follows from Fig.~\ref{fig:lambda_finite} an interesting complementarity between direct detection and collider constraints on the one hand, and searches for gamma-ray spectral features on the other hand: the LUX experiment and the LHC constrain the model for dark matter masses below $\simeq 100 - 200$ GeV (depending on the value of $\lambda$), while the Fermi-LAT and H.E.S.S. provide the strongest constraints for larger dark matter masses. Interestingly, some regions of the parameter space have been probed both by direct and indirect searches (and in some cases also by collider searches). The non observation of a dark matter signal then allows to more robustly exclude that part of the parameter space, in spite of the astrophysical uncertainties that plague the calculation of the direct and indirect detection rates.  Conversely, some regions of the parameter space will be probed in the next years both by the XENON1T experiment and by CTA, thus opening the exciting possibility of observing dark matter signals in more than one experiment. 

\section{Summary and Conclusions}
\label{sec:conclusions}

We have studied the generation of sharp gamma-ray spectral features in a toy model consisting in a scalar particle as dark matter candidate, that couples to a heavy exotic vector-like fermion and a Standard Model fermion via a Yukawa coupling.  More specifically, we have calculated the cross section for the processes generating gamma-ray lines at the one loop level and generating line-like features through the two-to-three annihilation into two Standard Model fermions with the associated emission of a gauge boson.  We have showed that the cross section for the two-to-three process is larger than for the loop annihilation when the fermionic mediator is degenerate in mass with the dark matter particle, the importance of the latter increasing as the ratio between the mediator mass and the dark matter mass becomes larger and larger. 

We have also calculated the expected intensity of the sharp spectral features for a dark matter population produced via the freeze-out mechanism. In this model, the annihilation cross section into a fermion-antifermion pair proceeds in the d-wave, therefore reproducing the correct relic abundance via thermal production requires a rather large Yukawa coupling, which in turn translates into relatively intense indirect detection signals. In fact, we find that large parts of the parameter space are already excluded by the Fermi-LAT and H.E.S.S. searches for line-like features. We have also investigated  the limits on the model from other indirect dark matter searches, direct searches and collider searches, and we have discussed the complementarity of these limits with those from the non-observation of sharp features in the gamma-ray sky.

\section*{Note Added}

During the completion of this work, we learned about an analysis of gamma-ray spectral features in the scalar dark matter model \cite{Giacchino2014}. Their results agree with ours, in the aspects where our analyses
overlap.
\section*{Acknowledgments}
We are grateful to Federica Giacchino, Laura Lopez-Honorez and Michel Tytgat for communications and for sharing with us the results of \cite{Giacchino2014}. We also thank Celine Boehm and Stefan Vogl for useful discussions. TT acknowledges support from the European ITN project (FP7-PEOPLE-2011-ITN, PITN-GA-2011-289442-INVISIBLES). AI, MT and SW were partially supported by the DFG cluster of excellence ``Origin and Structure of the Universe'', and SW further acknowledges support from the TUM Graduate School and the Studienstiftung des Deutschen Volkes.

\newpage
\appendix
\section*{Appendix A}

In this appendix we provide the complete expressions for the form factors $\mathcal{A}_{\gamma \gamma}$ and $\mathcal{A}_{\gamma Z}$ (see Eq.~(\ref{eq:MGammaGamma}) and~(\ref{eq:MGammaZ}) for its definitions)\footnote{We have used FeynCalc~\cite{Mertig:1990an} and LoopTools~\cite{Hahn:1998yk} for parts of the computations.}. We work exclusively in the limit $v \rightarrow 0$, i.e. we only keep the (dominant) s-wave term of the annihilation cross sections.

For $\mathcal{A}_{\gamma \gamma}$, we find 
\begin{align}
\mathcal{A}_{\gamma\gamma} \, &= 2 + m_\chi^2  \,\bigg\{\nonumber\\
&\frac{1-\mu-\epsilon}{1+\mu-\epsilon}
\frac{2\epsilon}{\mu-\epsilon}
C_0\left(-m_\chi^2,m_\chi^2,0;m_f^2,m_\psi^2,m_f^2\right)\nonumber\\
+&\frac{1-\epsilon-\mu}{1+\epsilon-\mu}
\frac{2\mu}{\epsilon-\mu}
C_0\left(-m_\chi^2,m_\chi^2,0;m_\psi^2,m_f^2,m_\psi^2\right)\nonumber\\
+&\frac{4\epsilon\left(1-\epsilon\right)}{1+\mu-\epsilon}
C_0\left(4 m_\chi^2 ,0,0;m_f^2,m_f^2,m_f^2\right)\nonumber\\
+&\frac{4\mu\left(1-\mu\right)}{1+\epsilon-\mu}
C_0\left(4m_\chi^2 ,0,0;m_\psi^2,m_\psi^2,m_\psi^2\right)\bigg\}.
\label{eq:exact_gamma}
\end{align}

where $\epsilon = m_f^2/m_\chi^2$ and $\mu = m_\psi^2/m_\chi^2$.
$C_0$ is the scalar three-point Passarino-Veltman integral~\cite{Passarino:1978jh} defined by
\begin{align}
&C_0\left(p_1^2,\left(p_1-p_2\right)^2,p_2^2;m_1^2,m_2^2,m_3^2\right)
\nonumber\\
&=
\int\frac{d^d\ell}{i\pi^2}
\frac{1}{\ell^2-m_1^2}\frac{1}{\left(\ell+p_1\right)^2-m_2^2}
\frac{1}{\left(\ell+p_2\right)^2-m_3^2}.
\end{align}

As already mentioned in the main text, our result differs from the one reported in~\cite{Tulin:2012uq}. We have checked that our full expression of $i \mathcal{M}_{\gamma \gamma}$ satisfies the Ward identity, and we also cross-checked the imaginary part of $\mathcal{A}_{\gamma \gamma}$ with the one deduced from the optical theorem~\cite{Asano:2012zv}. In the limit $\epsilon\to0$, one can explicitly evaluate all Passarino-Veltman functions in eq.~(\ref{eq:exact_gamma}), leading to the result given in eq.~(\ref{eq:loop-f}).

For the $\chi\chi\to \gamma Z$ process, we find \newpage

\begin{widetext}
\begin{eqnarray}
\mathcal{A}_{\gamma Z}\!\!&\,=\,&\!\!
2-\frac{\xi}{4-\xi}B_0\left(m_Z^2;m_f^2,m_f^2\right)
-\frac{\xi}{4-\xi}B_0\left(m_Z^2;m_\psi^2,m_\psi^2\right)\nonumber\\
&&
+\frac{2\xi\left(1+\mu+\epsilon\right)}
{\left(4-\xi\right)\left(1+\mu-\epsilon\right)\left(1+\epsilon-\mu\right)}
\left[1-\frac{1-\mu+\epsilon}{1+\mu+\epsilon}\frac{\epsilon}{2}\right]
B_0\left(m_\chi^2;m_f^2,m_\psi^2\right)\nonumber\\
&&
-\frac{\epsilon}{1+\mu-\epsilon}\frac{\xi}{4-\xi}
B_0\left(4m_\chi^2;m_f^2,m_f^2\right)
-\frac{2\mu}{1+\epsilon-\mu}\frac{\xi}{4-\xi}
B_0\left(4m_\chi^2;m_\psi^2,m_\psi^2\right)\nonumber\\
&&
+m_\chi^2\left\{
\frac{\epsilon}{2}\frac{4-4\epsilon-\xi}{1+\mu-\epsilon}
C_0\left(m_Z^2,4m_\chi^2,0;m_f^2,m_f^2,m_f^2\right)
+\mu\frac{4-4\mu-\xi}{1+\epsilon-\mu}
C_0\left(m_Z^2,4m_\chi^2,0;m_\psi^2,m_\psi^2,m_\psi^2\right)\right.\nonumber\\
&&
+\frac{\epsilon}{2}\left[\frac{\left(4+\xi\right)\left(-2+2\mu+2\epsilon+\xi\right)}
{\left(1+\mu-\epsilon\right)\left(4\epsilon-4\mu+\xi\right)}
+\frac{1}{2}\frac{4-4\epsilon-\xi}{1+\mu-\epsilon}
\right]
C_0\left(-m_\chi^2+\frac{m_Z^2}{2},m_\chi^2,0;m_f^2,m_\psi^2,m_f^2\right)\nonumber\\
&&
+\frac{\mu}{2}\left[\frac{\left(4+\xi\right)\left(-2+2\epsilon+2\mu+\xi\right)}
{\left(1+\epsilon-\mu\right)\left(4\mu-4\epsilon+\xi\right)}
-\frac{8\epsilon}{4\mu-4\epsilon+\xi}
\right]
C_0\left(-m_\chi^2+\frac{m_Z^2}{2},m_\chi^2,0;m_\psi^2,m_f^2,m_\psi^2\right)\nonumber\\
&&
+\left[
\frac{2\xi\left(1+\mu\right)+\epsilon\left(4\mu-\xi\right)}{4\left(1+\mu-\epsilon\right)}
-\frac{4\left(1+\mu\right)}{4-\xi}+\frac{4\mu\left(1+\mu-2\epsilon\right)}{4\mu-4\epsilon+\xi}
\right]C_0\left(-m_\chi^2+\frac{m_Z^2}{2},m_\chi^2,m_Z^2;m_f^2,m_\psi^2,m_f^2\right)\nonumber\\
&&
+\left.\left[
\frac{2\mu\left(1-\mu+3\epsilon\right)+\xi\left(1+\epsilon\right)}
{2\left(1+\epsilon-\mu\right)}
-\frac{4\left(1+\mu+\epsilon\right)}{4-\xi}
+\frac{4\epsilon\left(1+3\mu+\epsilon\right)}{4\epsilon-4\mu+\xi}
\right]
C_0\left(-m_\chi^2+\frac{m_Z^2}{2},m_\chi^2,m_Z^2;m_\psi^2,m_f^2,m_\psi^2\right)
\right\},\nonumber\\
\end{eqnarray}
\end{widetext}
where $B_0$ is defined by 
\begin{equation}
B_0\left(p_1^2;m_1^2,m_2^2\right)=
\int\frac{d^d\ell}{i\pi^2}
\frac{1}{\ell^2-m_1^2}\frac{1}{\left(\ell+p_1\right)^2-m_2^2}.
\end{equation}

%%%%%%%%%%%%%%%%%%%%%%%%%%%%%%%%%%%%

\end{document}